\documentclass[reprint,twocolumn,superscriptaddress,longbibliography]{revtex4-2}

\usepackage[hidelinks]{hyperref} 
\usepackage{graphicx}
\usepackage{amsmath}
\usepackage{amssymb}
\usepackage{siunitx}
\usepackage{braket}

\DeclareGraphicsExtensions{.png,.jpg,.eps}
\usepackage{xcolor}
\hypersetup{
    colorlinks,
    linkcolor={red!50!black},
    citecolor={blue!80!black},
    urlcolor={blue!80!black}
}

\renewcommand{\bm}{\boldsymbol}

\def\m1{{^{-1}}}

\begin{document}

\author{Bastian Zinkl}
\affiliation{Institute for Theoretical Physics, ETH Zurich, 8093 Zurich, Switzerland}

\author{Manfred Sigrist}
\affiliation{Institute for Theoretical Physics, ETH Zurich, 8093 Zurich, Switzerland}

\title{Impurity induced magnetic ordering in Sr$_2$RuO$_4$}

\begin{abstract}

Ti substituting Ru in Sr$_2$RuO$_4$ in small concentrations induces incommensurate spin density wave order with a wave vector $\bm{Q} \simeq (2 \pi /3, 2 \pi /3)$ corresponding to the nesting vector of two out of three Fermi surface sheets. We consider a microscopic model for these two bands and analyze the correlation effects leading to magnetic order through non-magnetic Ti-doping. For this purpose we use a position dependent mean field approximation for the microscopic model and a phenomenological Ginzburg-Landau approach, which both deliver consistent results and allow us to examine the inhomogeneous magnetic order. Spin-orbit coupling additionally leads to spin currents around each impurity, which in combination with the magnetic polarization produce a charge current pattern. This is also discussed within a gauge field theory in both charge and spin channel. This spin-orbit coupling effect causes an interesting modification of the magnetic structure, if currents run through the system. Our findings allow a more detailed analysis of the experimental data for Sr$_{2}$Ru$_{1-x}$Ti$_{x}$O$_{4}$. In particular, we find that the available measurements are consistent with our theoretical predictions. 

\end{abstract}

\date{\today}

\maketitle



\section{Introduction}
Motivated by the discovery of unconventional superconductivity, Sr$_2$RuO$_4$ has been studied extensively for well over two decades due to its many intriguing properties \cite{maeno1994superconductivity,maeno2001maeno,mackenzie2003superconductivity,kallin2012chiral}. However, the nature of the superconducting phase as well as its underlying pairing mechanism are still under debate, in particular in view of recent experiments putting the chiral $p$-wave phase as the previously most prominent candidate for the pairing symmetry into question \cite{chronister2020evidence,ishida2020reduction,mackenzie2020}. Due to its strongly correlated Fermi-liquid properties in the normal state Sr$_2$RuO$_4$ with its single-layer perovskite structure has been considered as a two-dimensional analog of $^3$He~\cite{rice1995,Bergemann2003quasi}. Moreover, it belongs to a Ruddelson-Popper series whose end member is the ferromagnetic compound SrRuO$_3$. Consequently, the pairing was anticipated to be mediated by ferromagnetic fluctuations. Interestingly, ferromagnetism is not the dominant part of the spin correlations in Sr$_2$RuO$_4$. Instead, incommensurate (IC) correlations have been experimentally observed at the wave vector $\bm{Q} \simeq (2 \pi /3, 2 \pi /3)$ (lattice constant taken as unit of length)~\cite{Mori:1999,Servant:2000,Braden:2002,Servant:2002,Braden:2004,Nagata:2004,Iida:2011}, very consistent with theoretical predictions~\cite{Mazin:1999,Nomura:2000magn,Takimoto:2000,Ng:2000,Morr:2001}.\par 

The IC wave vector is attributed to the nesting of two out of three Fermi surface (FS) sheets, which are dominated by the $4d$-$t_{2g}$ orbitals of the Ru$^{4+}$ ions. The $d_{xy}$ orbital yields the two-dimensional $\gamma$ band, while the $d_{xz}$ and $d_{yz}$ orbitals give rise to the $\alpha$ and $\beta$ bands, which incorporate quasi-one-dimensional features 
\cite{oguchi1995electronic,Damascelli:2000}. Given their one-dimensional character favoring Fermi surface nesting, these latter two are responsible for the IC correlation \cite{mazin1997ferromagnetic,Ng:2000}.\par 

Sr$_2$RuO$_4$ falls barely short of establishing magnetic long-range order originating from the strong IC correlations. It was, however, found that replacing some Ru$^{4+}$ by non-magnetic Ti$^{4+}$ ($3d^0$) induces magnetic order already at rather low Ti concentrations, $x > 0.025$ in Sr$_2$Ru$_{1-x}$Ti$_x$O$_4$ \cite{Minakata:2001}. Inelastic neutron scattering reveals a transition to a phase corresponding to a two-dimensional IC spin density wave state, whose moments are oriented along the $c$-axis and are modulated corresponding to wave vectors related to the Fermi surface nesting \cite{Braden:2002Ti,Iida:2012}. Naturally, Ti-doping suppresses the unconventional superconducting phase rather quickly even before magnetic order is established. Therefore there is no direct interference between the two ordered phases, although understanding magnetic correlations in Sr$_2$Ru$_{1-x}$Ti$_x$O$_4$ could shed light on the magnetic fluctuations of the pure compound. 

\begin{figure}[t!]
 \centering
	\includegraphics[width=0.80\columnwidth]{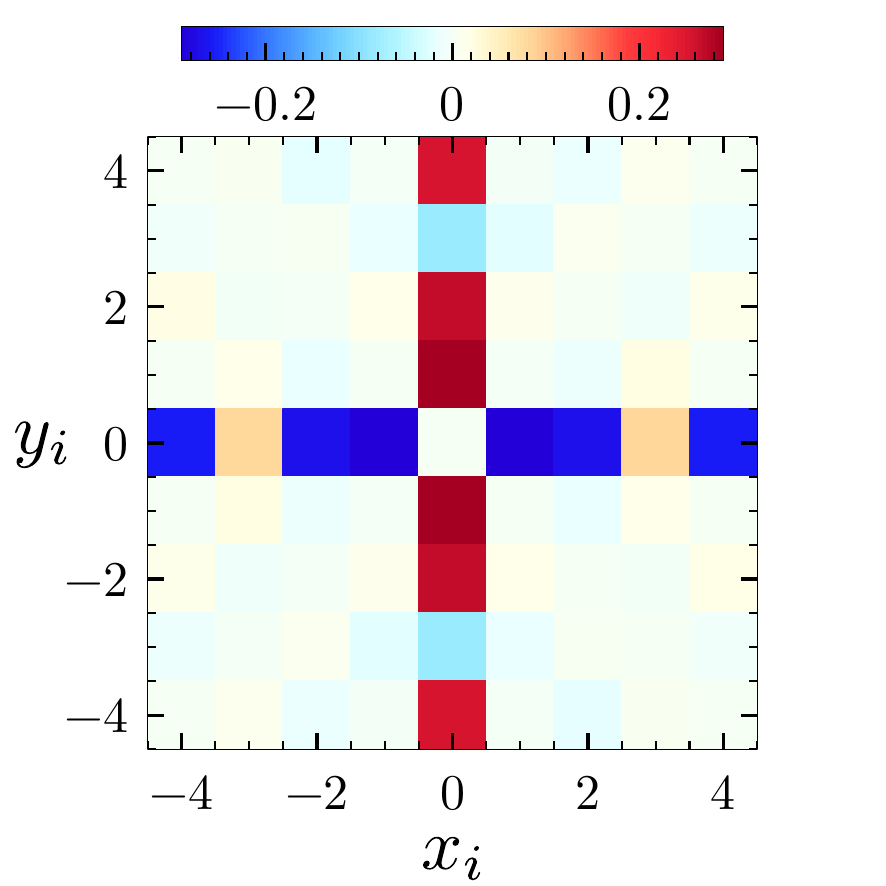}
	\caption{Spin pattern obtained through the dominant eigenvalue of the spin susceptibility matrix $ \chi_{zz} (\bm{r},\bm{r}') $ around a non-magnetic impurity in the microscopic two-orbital lattice model for the $ \alpha $-$\beta$ sheets of the Fermi surface. The impurity is located in the center and the system size is $9 \times 9$. Note that the spin susceptibility is computed in the non-interacting system, i.e. $ U=K=J=0 $ in the Hamiltonian [Eq.~\eqref{hamil-complet}].}
	\label{fig:m-cross}
\end{figure}

Interestingly, non-magnetic impurities have been found to cause magnetic order also in insulating spin systems with antiferromagnetic short-range correlations and a spin excitation gap. In particular, a series of 
cuprate systems show this feature upon replacing some Cu-ions by Zn impurities \cite{Zn-CuGeO3,Zn-NMR}. This type of systems has been investigated theoretically quite extensively in the context of the concept {\it order by disorder} and the modification of magnetic quantum phase transitions by impurities \cite{Matsumoto-2001,Wessel-2001,Laflorencie-2009}. The nucleation of a pattern of magnetic order around a non-magnetic impurity observed in these model systems is rather similar to our metallic case. However, there are also obvious differences concerning the coupling of several impurities and the fact that our system is metallic. A recent, prominent case of a metallic system close to a magnetic quantum phase transition is FeSe, which shows antiferromagnetic order in films epitaxially grown on SrTiO$_3$, while bulk FeSe does not exhibit magnetic order \cite{PhysRevLett.120.097001}. It has been suggested that non-magnetic defects and impurities could be responsible for this behavior, whereby in this case superconductivity could play a role \cite{PhysRevLett.124.117001,PhysRevB.99.014509,PhysRevB.96.094504}.  

The aim of our study is to explore the formation of magnetic order through non-magnetic impurities from a microscopic point of view. Since the magnetic state is clearly governed by the Fermi surface nesting of the $\alpha$ and $\beta$ sheets, we will neglect the third ($\gamma $) sheet for simplicity in the following discussion. From the microscopic model of the $\alpha$-$\beta$-bands including spin-orbit coupling we develop a mean field theory for a spatially dependent magnetization of the two involved $4d$-orbitals and determine the pattern of magnetic order around a single impurity. We find that the dominant magnetic correlation in the pure system is connected with the four IC wave vectors $ \bm{Q}  \simeq ( \pm 2 \pi /3, 0)$ and $(0,  \pm 2 \pi /3) $, which, if strong enough, would lead to a spin density wave state based on these wave vectors. In this study, our focus lies on establishing the qualitative structure of the magnetization pattern induced by impurities, meaning we do not attempt to give a quantitative discussion. 

A simple probe of the magnetic response in the presence of an impurity can be obtained by considering the spin susceptibility $ \chi_{zz} (\bm{r},\bm{r}') $ in real space on a square lattice using the electron structure of the $ \alpha $-$\beta$-bands without interactions. We observe that the dominant spin correlation yields a cross like pattern with the impurity at the center, as displayed in Fig.~\ref{fig:m-cross}.  Each beam of the cross can be attributed to one of the two orbitals, $4d_{yz}$ and $4d_{zx}$, and the spin polarization is modulated by the corresponding nesting vectors $ \bm{Q}^{(x)} \simeq ( 2 \pi / 3 ,0 ) $ and $ \bm{Q}^{(y)} \simeq (0, 2 \pi / 3 ) $ in $x$- and $y$-direction, respectively. Moreover, the sign is reversed on the two beams such that no net magnetic moment is associated with the impurity. Indeed, we will show below that this is the basic structure, which appears spontaneously due to the interaction around the impurities.

In the following, we will analyze the magnetic structure induced by non-magnetic impurities based on a microscopic model of the $\alpha$-$\beta$-bands including the effects of spin-orbit coupling [Sec.~\ref{sec:origin}]. A first attempt using a theory based on averaged impurity scattering does not indicate any inclination of the system towards magnetic order. Thus, we turn towards a mean field approach with spatial resolution, which allows us to show that a non-vanishing spin polarization appears around the impurity due to the repulsive onsite potential. In order to further analyze our findings we consider also the multi impurity situation, which we examine using a phenomenological Ginzburg-Landau formulation [Sec.~\ref{sec:GLsection}]. The Ginzburg-Landau theory is subsequently supported by a field theory derived from the microscopic model. The different approaches fall very much in line and provide a clear picture of the structure of the impurity induced magnetic order. Extending the field theory into a gauge field theory of the charge and spin degrees of freedom reveals several more features of the system, such as (impurity) potential induced spin currents and the connection between spin polarization and charge currents [Sec.~\ref{sec:EFTsection}]. These phenomena may allow for interesting probes of the magnetic properties of such a system.

\section{Magnetism of the two-orbital system} \label{sec:origin}

After introducing the basic two-band model we analyze the magnetic properties first using a formalism based on configuration-averaged impurity scattering and then by focusing on a local mean field theory for a single impurity.

\subsection{The model Hamiltonian} \label{sec:modelHamil}

\begin{figure}[t!]
 \centering
	\includegraphics[width=0.80\columnwidth]{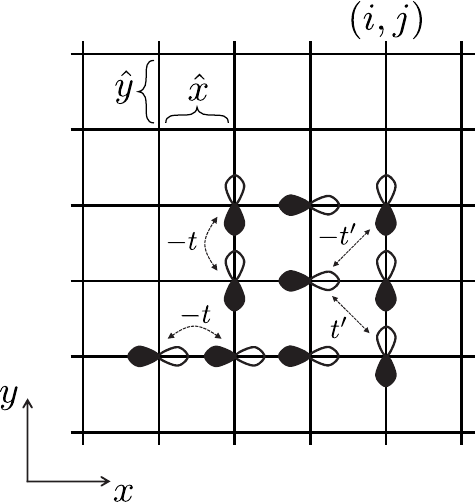}
	\caption{
		Illustration of all possible square lattice hopping terms for the $x$ and $y$ orbitals. The intra-orbital (inter-orbital) hopping matrix element is denoted by $t$ ($t'$). 
	}
	\label{fig:hopping}
\end{figure}

Our model Hamiltonian describes the $\alpha$- and $\beta$-band of Sr$_2$RuO$_4$, which are derived from the $4d$-$t_{2g} $-orbitals $ d_{xz} $ and $ d_{yz} $, labelled as '$x$' and '$y$' in the following. It includes nearest-neighbor (NN) intra-orbital and next-nearest neighbor (NNN) inter-orbital hopping corresponding to the pattern depicted in Fig.~\ref{fig:hopping}, onsite spin-orbit coupling and Coulomb interactions.  
We split the Hamiltonian into four parts, 
\begin{align}
	H = H_{b} + H_{SO} + H_{U} + H_{\text{imp}}, 
	\label{hamil-complet}
\end{align}
where $ H_b $ describes the tight-binding bands of the two orbitals, $ H_{SO}$ corresponds to the spin-orbit coupling, $ H_U $ to the onsite Coulomb interaction between electrons and $ H_{\text{imp}} $ is the impurity potential. These terms are given by \begin{align}
	H_{b} =& -t \sum_{\bm{r},s} \left( a_{\bm{r} x s}^{\dagger} a_{\bm{r}+\hat{\bm{x}} x s} + a_{\bm{r} y s}^{\dagger} a_{\bm{r}+\hat{\bm{y}} y s}\right) \nonumber \\
&-t' \sum_{\bm{r},\alpha,s} \left( a_{\bm{r} {\alpha} s}^{\dagger} a_{\bm{r}+\hat{\bm{x}}+\hat{\bm{y}} {\bar{\alpha}} s} - a_{\bm{r} {\alpha} s}^{\dagger} a_{\bm{r}+\hat{\bm{x}}-\hat{\bm{y}}  {\bar{\alpha}} s} \right) \nonumber \\
	 &- \mu\sum_{\bm{r},\alpha,s} n_{\bm{r} {\alpha} s} + h.c, \label{hb} \\[2mm]
	H_{SO} = &- \lambda \sum_{\bm{r},s} i s a_{\bm{r} x s}^{\dagger} a_{\bm{r} y s} + h.c. , \label{hso} \\[2mm]
H_{\text{imp}} =& \; V \sum_{\bm{R}_j,\alpha,s} 
a_{\bm{R}_j {\alpha} s}^{\dagger} a_{\bm{R}_j {\alpha} s}  \label{eqn:hamiltonian} \\[2mm]
H_{U} = & \; U \sum_{\bm{r},\alpha} n_{\bm{r} {\alpha} \uparrow} n_{\bm{r}{\alpha} \downarrow} + K \sum_{\bm{r},\alpha} n_{\bm{r} {\alpha} \uparrow} n_{\bm{r} {\bar{\alpha}} \downarrow} \nonumber \\
&+ (K - J ) \sum_{\bm{r},s} n_{\bm{r} {x} s} n_{\bm{r} {y} s}, \label{eqn:hamil-Coul}	
\end{align}
where $a_{\bm{r} {\alpha} s}^{\dagger}$ ($a_{\bm{r} {\alpha} s}$) is the creation (annihilation) operator for electrons on site $ \bm{r}$, in orbital ${\alpha} = \{ x, y \} $ and with spin $s = \{1 \ \widehat{=} \uparrow, -1 \ \widehat{=} \downarrow \}$ and $ n_{\bm{r}, \alpha, s} = a_{\bm{r}, \alpha,s}^{\dag}  a_{\bm{r}, \alpha,s} $ is the density operator. The vectors $ \hat{\bm{x}} $ and $ \hat{\bm{y}} $ are primitive lattice vectors of the square lattice in $x$- and $y$-direction, respectively (see Fig.~\ref{fig:hopping}).We use the notation ${\bar{\alpha}} = \{ y, x \}$ and restrict the lattice to $0 \leq i,j < N$, where $N$ is the linear system size. The chemical potential and the spin-orbit coupling constant are denoted by $\mu$ and $\lambda$, respectively. Impurities at the positions $ \bm{R}_j $ exhibit an onsite repulsive potential $ V > 0$, which is assumed to be much larger than the band width.  
The interaction term includes intra-orbital $U$, inter-orbital $K$ and Hund's rule coupling $J$. For the last one we neglected the pair hopping, for simplicity, without changing our results qualitatively. We also assume the standard relation between interaction strengths, $U = K+ 2 J $.\par

\begin{figure}[t!]
 \centering
	\includegraphics[width=0.90\columnwidth]{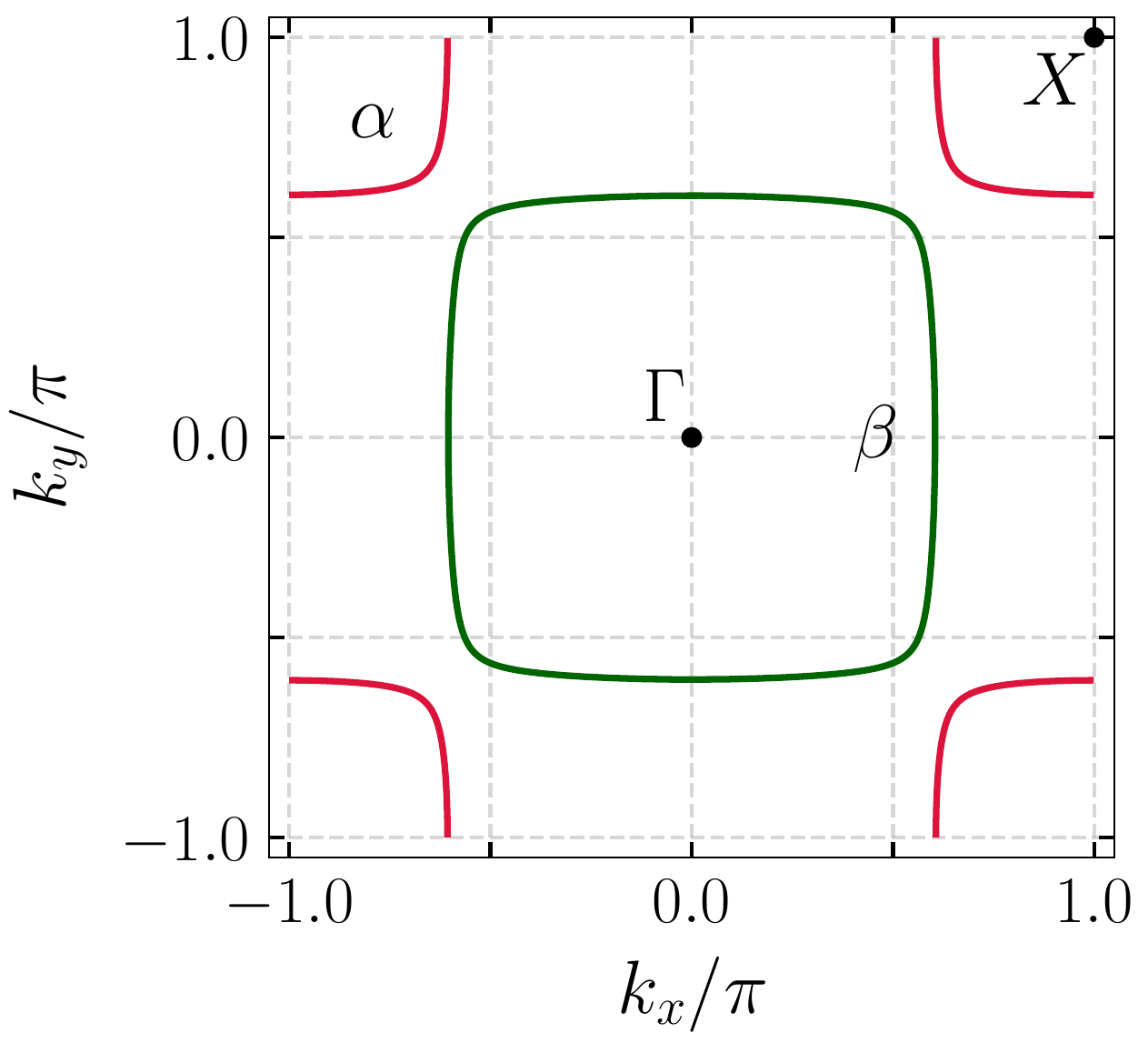}
	\caption{
	Fermi surfaces of the hole-like $\alpha$ (red) and electron-like $\beta$ (green) sheets for $\mu = 0.065$, reproducing 
	approximately the band structure of these two bands of Sr$_2$RuO$_4$.
	}
	\label{fig:FS}
\end{figure}

We can rewrite the Hamiltonian in $ \bm{k} $ space using the spinor notation $ C_{\bm{k}, s} = (c_{\bm{k} x s}, c_{\bm{k} y s})^{T} $, where
\begin{equation}
a_{\bm{r} \alpha s} = \frac{1}{N} \sum_{\bm{k}} c_{\bm{k} \alpha s} e^{i \bm{k} \cdot \bm{r}} .
\end{equation}
 Thus, the single-particle part combining $ H_b $, $ H_{SO} $ and $ H_{\text{imp}} $ yields
 \begin{equation}
 H_0 = \sum_{\bm{k},\bm{q},s} C_{\bm{k}_-, s}^{\dag} \left( \hat{H}_{\bm{k},s} \delta_{\bm{q},\bm{0}} +  V_{-\bm{q}} \hat{1} \right) C_{\bm{k}_+,s} ,
 \label{H0}
 \end{equation}
 with $ \bm{k}_{\pm} = \bm{k} \pm \bm{q}/2 $ and
 \begin{equation}
    \hat{H}_{\bm{k},s} = \begin{pmatrix}
    	\xi_{x,\bm{k}} & \gamma_{\bm{k}} - i\lambda s \\[2mm]
    	\gamma_{\bm{k}} + i\lambda s & \xi_{y,\bm{k}}
    \end{pmatrix} ,
    \label{H-matrix}
\end{equation}
where $\xi_{\alpha,\bm{k}} = -2t\cos k_{\alpha} - \mu$ and $\gamma_{\bm{k}} = 4t' \sin k_x \sin k_y$ and $ V_{-\bm{q}} = \sum_{\bm{R}_j} V e^{i \bm{q} \cdot \bm{R}_j} $ ($s=\pm 1 $). The interaction part can be written as
\begin{align}
H_{U} 
& = \sum_{\bm{q}, \alpha,\alpha'} \left\{ \rho_{\bm{q} \alpha} K_{\alpha \alpha'} \rho_{-\bm{q} \alpha'} - m_{\bm{q} \alpha} U_{\alpha \alpha'} m_{-\bm{q} \alpha'} \right\} \label{HU},
\end{align}
where 
\begin{equation}
\rho_{\bm{q} \alpha} = \frac{1}{N^2} \sum_{\bm{k},s} c_{\bm{k}_+ \alpha s}^{\dag} c_{\bm{k}_- \alpha s} 
\label{rho-q}
\end{equation}
and
\begin{equation}
m_{\bm{q} \alpha} = \frac{1}{2N^2} \sum_{\bm{k},s} s c_{\bm{k}_+ \alpha s}^{\dag} c_{\bm{k}_- \alpha s} .
\end{equation}
The interaction matrices in orbital space are given by
\begin{equation}
\hat{K} = \frac{1}{4} \begin{pmatrix} U & 2K-J \\ 2K-J & U \end{pmatrix} 
\end{equation}
and
\begin{equation}
\hat{U} = \begin{pmatrix} U & J \\ J& U \end{pmatrix} .
\end{equation}

Ignoring impurities we obtain the two-dimensional band structure shown in Fig.~\ref{fig:FS} from $H_0$, where the first Brillouin zone (BZ) is given by $k_x, k_y \in [-\pi, \pi]$ with the lattice constant $a=1$. In the following we take the NN hopping matrix element $ t $ as the unit of energy and choose $ t'=0.1 t $ and $ \lambda = 0.1 t $. The chemical potential is fixed to keep the electron density at $n=8/3$. Since the band structure is dominated by the NN intra-orbital hopping, the band structure retains a strong one-dimensional character leading to pronounced nesting wave vectors $\bm{Q} \simeq (2 \pi /3, 0)$ and $  (0, 2 \pi /3)$.

\subsection{Self-consistent T-matrix approximation}\label{t-matrix}
We first discuss the effect of Ti-impurities by considering configuration-averaged impurity scattering based on the self-consistent T-matrix approximation. This scheme uses the Green's function formalism and allows us to calculate straightforwardly the spin susceptibility, 
\begin{align}
    \chi_{zz}^{\alpha \alpha'} (\bm{q}, \tau) = - \braket{T_{\tau} S^{z}_{\alpha,\bm{q}}(\tau) S^{z}_{\alpha',-\bm{q}}(\tau)},
\end{align}
where $S^{z}_{\alpha,\bm{q}} = (1/N) \sum_{\bm{k}, s, s'} c_{\bm{k+q} {\alpha} s}^{\dagger} \sigma^{z}_{s s'} c_{\bm{k} {\alpha} s'} $ with $ c_{\bm{k} {\alpha} s}^{\dagger} $ and $ c_{\bm{k} {\alpha} s} $ denoting the creation and annihilation operators, respectively, for electrons of momentum $ \bm{k} $ in orbital $ \alpha $ with spin $s$. 
We restrict ourselves to the $z$-axis spin polarization because spin-orbit coupling yields an anisotropy, which boosts the $z$ axis component compared to the inplane polarizability for $ \bm{q} \approx \bm{Q} $, which is most relevant for the magnetic correlations in our discussion \cite{Ng:2000}. 

The static spin susceptibility can be expressed by means of Green's functions, $G_{ s s'}^{\alpha \alpha'}(\bm{k}, \tau) = -i \braket{T_{\tau} c_{\bm{k} {\alpha} s}(\tau)  c_{\bm{k} {\alpha'} s'}^{\dagger}(0)}$, as
\begin{align}
    \chi_{zz}^{\alpha \alpha'} (\bm{q}) = &\frac{1}{4 \beta} \sum_{s_1, \dots, s_4, n} G_{s_2 s_3}^{\alpha \alpha'} (\bm{k}, i\omega_n) \nonumber \\[2mm]
    &\times G_{s_4 s_1}^{\alpha' \alpha} (\bm{k+q}, i\omega_n) \sigma_{s_1 s_2}^{z} \sigma_{s_3 s_4}^{z} ,\label{eqn:spinsusc} 
\end{align}
with the Fermionic (Bosonic) Matsubara frequencies $ \omega_n = (2n+1) \pi /\beta $ ($ \nu_l = 2l \pi/\beta $). We define the susceptibility of the orbital $\alpha$ as $\chi_{zz}^{\alpha} (\bm{q})= \sum_{\alpha'} \chi_{zz}^{\alpha \alpha'} (\bm{q})$. For the impurity-free system we use the bare Green's function given by
\begin{align}
    \hat{G}_{0,ss'}(\bm{k}, i\omega_n) = \frac{\delta_{ss'}}{\mathcal{D}_{\bm{k}}} \begin{pmatrix}
    	i\omega_n - \xi_{x,\bm{k}} & \gamma_{\bm{k}} - i\lambda s \\[2mm]
    	\gamma_{\bm{k}} + i\lambda s & i\omega_n - \xi_{y,\bm{k}}
    \end{pmatrix}, 
    \end{align}
which is derived straightforwardly from the equation,
\begin{align}
    \left( i \omega_n \hat{1} - \hat{H}_{\bm{k},s} \right) \hat{G}_{0,s}(\bm{k}, i\omega_n) = \hat{\sigma}_0,
\end{align}
with $ \hat{H}_{\bm{k},s} $ of Eq.~(\ref{H-matrix}). Because the bare Green's function is diagonal in the spin index, we use the
shorter notation $  \hat{G}_{0,s}(\bm{k}, i\omega_n) $ with a single spin index.

In line with the impurity potential introduced in the Hamiltonian [Eq.~\eqref{eqn:hamiltonian}] we write the intra-orbital single-impurity scattering term as
\begin{align}
    H_{\text{imp}}' = \sum_{\bm{k}, \bm{k'}, \alpha, s} V_{\bm{k}, \bm{k'}} c_{\bm{k} {\alpha} s}^{\dagger} c_{\bm{k} {\alpha} s},
\end{align}
with $ V_{\bm{k}, \bm{k'}} = V $, which corresponds to $s$-wave scattering. The impurity scattering leads to the renormalized Green's function $\hat{G}_{ss'} (\bm{k}, i\omega_n)$ defined through the Dyson equation, 
\begin{align}
    \hat{G}_{ss'}^{-1}(\bm{k}, i\omega_n) =  \delta_{ss'} \{ \hat{G}^{-1}_{0,s}(\bm{k}, i\omega_n) - \hat{\Sigma}_s(\bm{k} , i\omega_n) \}, \label{eqn:Dyson}
\end{align}
where $\hat{\Sigma}(\bm{k}, i\omega_n)$ denotes the self-energy.  Note that the Green's function is diagonal in the spin indices.

To account for a large impurity potential we employ the T-matrix approach, which includes multiple scatterings at the impurity  \cite{hirschfeld1986resonant, hirschfeld1988consequences, balatsky2006impurity}. 
Due to the $s$-wave character of the scattering the momentum dependence drops out of the self-energy, which is directly connected with the T-matrix,  $\hat{T}(i \omega_n)$, through 
\begin{align}
    \hat{\Sigma}_s(i\omega_n) = n_{\rm imp}  \hat{T}_s(i \omega_n).
\end{align}
This relation is only valid for small impurity concentrations $ n_{\rm imp} $, where interference effects between scattering events at several impurities can be neglected. 

Inserting it in Eq.~\eqref{eqn:Dyson} we determine $\hat{T}(i \omega_n)$ self-consistently, 
\begin{equation}
\hat{T}_s(i \omega_n) = V + \sum_{\bm{k}} V  \hat{G}_{ss}^{-1}(\bm{k}, i\omega_n) \hat{T}_s(i \omega_n) .
\end{equation}
\begin{figure}[t!]
 \centering
			\includegraphics[width=\columnwidth]{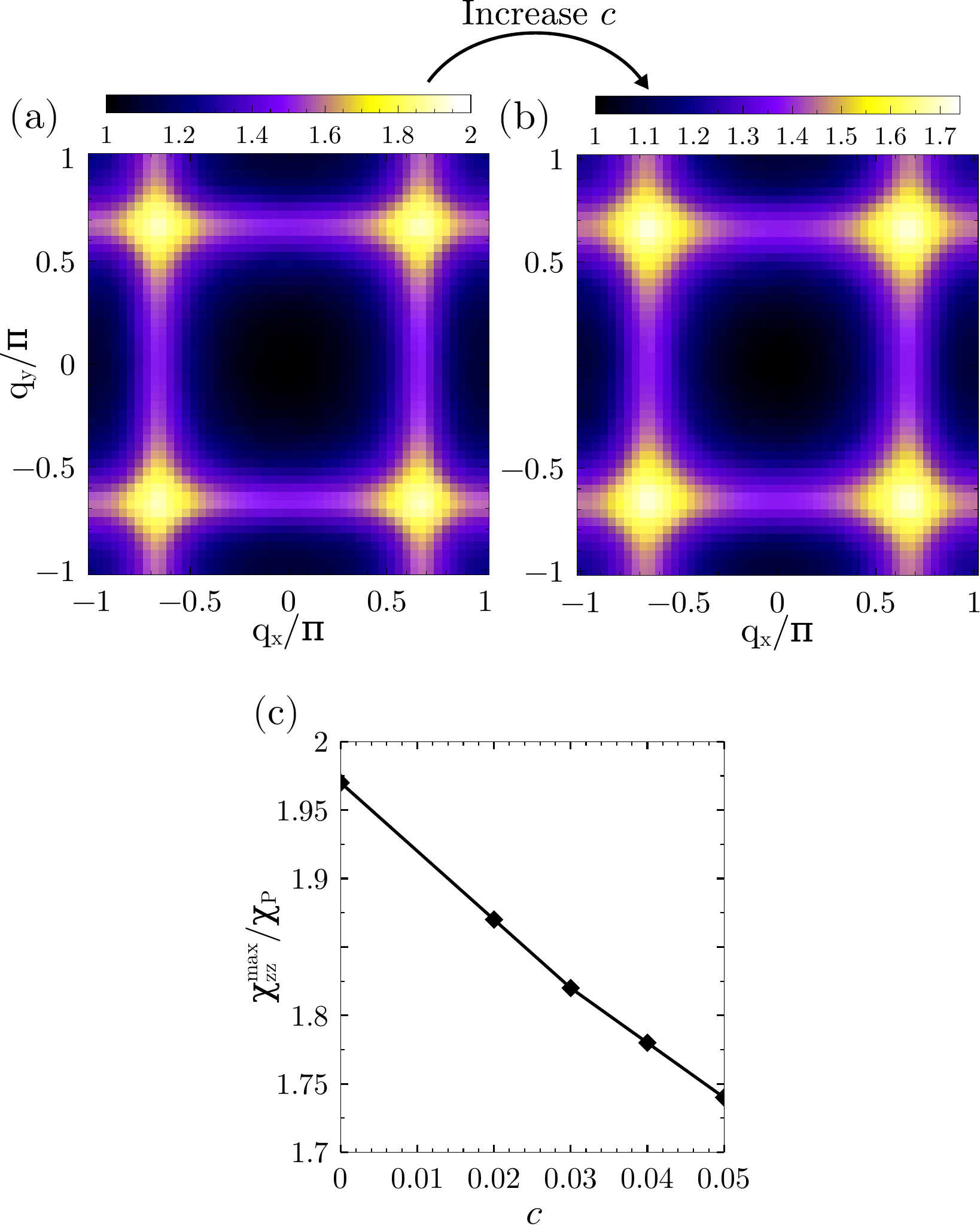}
\caption{Static spin susceptibility $\chi_{zz}^{(\alpha+\beta)}$ for (a) no impurities and (b) a finite concentration $ n_{\rm imp} = 0.05 $ of impurities in the non-interacting system ($U=K=J=0$). The susceptibilities are normalized with the Pauli susceptibility, $\chi_P$. In (c) the peak maximum of the susceptibility is plotted against the impurity concentration. 
}
\label{fig:tmatrixsusc}
\end{figure}

In Fig.~\ref{fig:tmatrixsusc} the static spin susceptibility $\chi_{zz}^{(x+ y)}(\bm{q})$, summed over both orbitals, is depicted for the clean (a) and the impurity-doped (b) system. In both cases we observe pronounced peaks at the four wave vectors $ \bm{Q} = (\pm 1, \pm 1) Q_0 $ with $ Q_0 \approx 2 \pi /3 $, which is not shifted noticeably by disorder. This is obviously a feature originating from the Fermi surface nesting as mentioned above. 
Impurity scattering introduces, however, a decrease in quasiparticle life time, which broadens the peaks while simultaneously reducing their height. This behavior is systematic as can be seen in Fig.~\ref{fig:tmatrixsusc} (c). Within this approach we cannot explain the observed magnetic ordering due to Ti-doping. On the contrary, impurity scattering yields apparently a weakening of magnetic correlations. This is a shortcoming of the averaging procedure we use here, which leaves the system formally translationally invariant. The magnetic order we will describe below, however, is not at all homogeneous, but strongly modulated in the system. Thus, a different approach is necessary to analyze the magnetic instability.

\subsection{Local mean field approximation} \label{sec:impindpol}

In contrast to the spatial average of the previous section, we analyze now the magnetic properties around a single impurity using a local mean field approach. 
For this purpose we decouple the interaction term $ H_U $ in our Hamiltonian [Eq.~\eqref{eqn:hamil-Coul}] by means of a Hartree-Fock mean field approximation \cite{bruus2004many},
\begin{align}
n_{\bm{r} {\alpha} s} =  \braket{n_{\bm{r} {\alpha} s}} + (n_{\bm{r} {\alpha} s}  -  \braket{n_{\bm{r} {\alpha} s}} )
\end{align}
where $ \braket{n_{\bm{r} {\alpha} s}} $ denotes the expectation value for the ground state. The resulting single-particle Hamiltonian is then solved self-consistently for the spatially resolved expectation values $\braket{n_{\bm{r}{\alpha} s}}$, which yield the local spin polarization, 
\begin{align}
    m_{\bm{r} {\alpha}} = \frac{1}{2} \left( \braket{n_{\bm{r} {\alpha} \uparrow}} - \braket{n_{\bm{r} {\alpha} \downarrow}} \right).
\end{align}
As mentioned above, the preference for $z$-axis orientation of the magnetic moments is caused by spin-orbit coupling. 

For our numerical calculation we choose the finite size system of a $ N \times N $-lattice with $ N =21 $ and a single impurity in the center. The use of periodic boundary conditions yields effectively a regular lattice of impurities, which are $ N $ lattice constants apart. The fact that $ N $ can be divided by 3 makes the cells commensurate with the nesting vector $ Q_0 = 2 \pi /3 $. In order to reduce other finite-size effects we average over twisted boundary conditions in the evaluation of the self-consistent equation, introducing phase factors to the hopping matrix elements through the boundaries. 

\begin{figure}[t!]
 \centering
	\includegraphics[width=0.99\columnwidth]{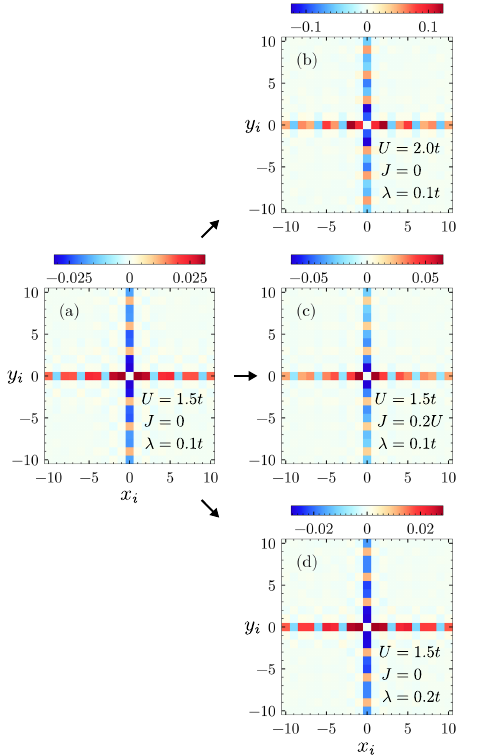}
	\caption{
		Spin polarization $(m_{\bm{r}_i, x} + m_{\bm{r}_i y})$ around a single impurity located at the center for different parameter values of $U$, $J$ and $ \lambda $ for a square of $21 \times 21 $ lattice sites: (a) shows the result of the basic parameter set; (b) the Coulomb repulsion $U$ is increased; (c) the Hund's rule coupling $ J $ is turned on; (d) the spin orbit coupling $ \lambda $ is increased. For the comparison of the magnitude of the spin polarization note that the color scale of each plot is different.	
	}
	\label{fig:spinpol}
\end{figure}

Initially we use the following model parameters for the band structure: $ t'= \lambda = 0.1t $, and for the Coulomb interaction $ U=1.5 t $ with $ J = 0 $ (keeping $ U=K+2J$). The impurity potential $V$ is several times the band width. The basic result displayed in Fig.~\ref{fig:spinpol} (a) shows a cross-shaped pattern of the mean field spin polarization with the impurity position at the center. The narrow beam along the $x$-axis ($y$-axis) is strongly dominated by the $ x $-($y$-)orbital. The pattern has a basic oscillatory modulation with the wave vector $ Q_0 $ in both directions and decays rather rapidly. The spin polarization has opposite signs for the two beam directions. Hence, its symmetry follows the irreducible representation $ B_1 $ of the point group $ C_{4v} $ of the square lattice with the basis function $ \Phi_{B_1} (\bm{r}) = x^2 - y^2$. This means that there is no net magnetic moment associated with the impurity.  
This basic finding is entirely in line with the structure obtained from the analysis of the local spin susceptibility around an impurity, discussed in the introduction [Fig.~\ref{fig:m-cross}].

To probe the influence of the different ingredients of our model on the structure, we vary some of the model parameters. The inter-orbital hopping $ t'$ and the spin-orbit coupling $ \lambda $ are essential for the opposite sign of the magnetization in $x$- and $y$-direction ($ B_1 $ symmetry). Neglecting the two, $ t' = \lambda = 0 $, leads to the degeneracy with the structure where both beams have the same sign (belonging to the irreducible presentation $ A_1 $). Increasing the Coulomb repulsion $ U $ strengthens the spin polarization keeping the same modulation [Fig.~\ref{fig:spinpol} (b)] (note the color scales given for each figure). Turning on the Hund's coupling raises the spin polarization as well [Fig.~\ref{fig:spinpol} (c)]. On the other hand, increasing the spin-orbit coupling reduces the spin polarization [Fig.~\ref{fig:spinpol} (d)], whereby the beams gain in width, a feature which is not easy to observe in our figures, but has been tested in our numerical results. The mechanism lies in the transfer of the spin polarization to the orbital polarization through spin-orbit coupling. The angular momentum operator,  $ \langle L_z \rangle \propto i \langle a_{\bm{r} xs}^{\dag}  a_{\bm{r}ys} - a_{\bm{r}ys}^{\dag}  a_{\bm{r}xs} \rangle $, involves both orbitals. Thus, on the beam along the $x$-axis the spin polarization of the $x$-orbital is shifted to the $ y $-orbital, which has preference to extend along the $y$-axis. This shift yields a reduction of the spin polarization directly on the beam axis. 

\section{Phenomenological approach for a system with many impurities} \label{sec:GLsection}
The mean field theory of the previous section considered the spin polarization induced by a single impurity. The fact that we used periodic (twisted) boundary conditions for a finite system led effectively to a regular lattice of impurities, which are rather far apart from each other and, thus, weakly coupled. In this section we would like to examine the interplay of impurities at nearby locations and the effect on the magnetization pattern. While this can, in principle, also be dealt with in our previous mean field setup, a phenomenological approach is here more efficient. We formulate a Ginzburg-Landau theory, which we compare with an effective field theory derived from our microscopic model in the subsequent section. 

\subsection{Ginzburg-Landau theory}\label{GL-Theory}

As in the mean field approach we consider the $z$-axis spin polarization in each orbital ($x$ and $y$) as a position dependent order parameter, $ m_x (\bm{r}) $ and $ m_y  (\bm{r}) $. These two order parameter components can be combined to belong to two irreducible representations of the square lattice point group $ C_{4h} $, 
\begin{align}
	&A_{1}: m_A = m_x + m_y, \\[2mm]
	&B_{1}: m_B= m_x - m_y,
\end{align}
where $m_x$ and $ m_y $ can be considered to transform like the basis functions $ x^2$ and $y^2$, respectively. Following Landau's scheme one could construct the Ginzburg-Landau theory based on the two order parameters $ m_A $ and $ m_B $, describing the spontaneous symmetry breaking in one of the two representations. However, it is actually more advantageous to formulate the free energy expansion with $ m_x $ and $ m_y$. Eventually, multiple impurities will reduce the point group symmetry and lead to a coupling of the order parameters of the two representations. 

The scalar free energy functional is given by
\begin{equation}
{\cal F} [m_x , m_y] := \int d^2r f(m_x,m_y),
\end{equation}
with  
\begin{align}
&f = a(T) (m_x^2 + m_y^2) + 2 c m_x m_y + b_1 m_A^4 + b_2m_B^4 + b_3 m_A^2 m_B^2  \nonumber \\[2mm] 
&+ K_1 \left[ - (\partial_x m_x)^2 + \kappa (\partial_x^2 m_x)^2 - (\partial_y m_y)^2 + \kappa (\partial_y^2 m_y)^2 \right] \nonumber  \\[2mm] 
&+ K_2 \left[ (\partial_x m_y)^2 + (\partial_y m_x)^2 \right] + K_3 \left[(\partial_x m_x) (\partial_x m_y) \right. \nonumber \\[2mm]  
& \left. +(\partial_y m_x) (\partial_y m_y) \right] + \gamma \sum_j (m_x^2 + m_y^2) \delta (\vec{r} - \vec{R}_j), \label{GL-FE}
\end{align}
where $ a(T) = a' (T-T_0) $ and $c $, $ b_{1,2,3} $ and $ K_{1,2,3} > 0 $. Note that we used the combinations $ m_A $ and $ m_B $ for the fourth order terms for compactness. 
The spatial modulation of the order parameter with the wave vector $ Q_0 $ is introduced by including higher order derivatives into the gradient terms where $ \kappa=1/2Q^2 $ with $ Q \approx Q_0 = 2 \pi/3$. The last term represents the effect of the impurities where the coupling $ \gamma < 0 $ facilitates the local nucleation of the order parameter. We ignore any internal structure of the impurity sites and approximate it by a $\delta$-function for simplicity. 

In order to analyze the nucleation of the impurity induced magnetic pattern we restrict ourselves to the linearized Ginzburg-Landau equation obtained through the variation of $ {\cal F} [m_x,m_y] $ with respect to the two order parameter components and by neglecting non-linear terms in $ m_{x,y} $, 

\begin{equation}
\sum_{\alpha'} A_{\alpha \alpha'} (\bm{q},T)    m_{\bm{q}}^{\alpha'} = -\gamma \sum_j e^{-i \bm{q} \cdot \bm{R}_j} \Pi_{\alpha,j},
\label{GL-q}
\end{equation}
which is written in $ \bm{q} $-space using the Fourier transformation 
\begin{equation}
m_{\alpha} (\bm{r}) = \int d^2q \; m_{\bm{q}}^{\alpha} e^{i \bm{q} \cdot \bm{r}} .
\label{m-Fourier}
\end{equation}
The symmetric matrix $ A_{\alpha \alpha'} $ in $\bm{q}$-space reads 
\begin{align}
A_{\alpha \alpha}(\bm{q},T)  &= a(T) + K_1 (\kappa q_{\alpha}^4 - q_{\alpha}^2) + K_2 q_{\bar{\alpha}}^2, \\[2mm]
A_{xy}(\bm{q}) & = A_{yx}(\bm{q}) = c + \frac{K_3}{2} (q_x^2 + q_y^2) 
\label{Aaa}
\end{align} 
and the right hand side of Eq.~\eqref{GL-q} uses 
\begin{align}
\Pi_{\alpha, j} &= \int d^2 q \ m_{\bm{q}}^{\alpha} e^{i \bm{q} \cdot \bm{R}_j} \label{Pi}. 
\end{align} 
The linearized GL equation (\ref{GL-q}) includes also the bulk instability. Setting $ \gamma =0 $, the bulk instability to an incommensurate spin density wave state occurs at the temperature where the first of the two eigenvalues 
of the $2 \times 2 $-matrix $ \hat{A} (\bm{q},T) $ changes sign for a certain $ \bm{q} $ ( $\approx \bm{Q}_0 $) from a positive to a negative value at a temperature $ T=T_N < T_0$. In Sr$_2$RuO$_4$ the bulk instability condition would lead to $ T_N < 0 $. 
We restrict ourselves to a temperature range $ T > T_N $, where for all $ \bm{q} $ the matrix $ \hat{A}(\bm{q},T) $ is strictly positive definite. 
We introduce $ \hat{\eta}(\bm{q},T) =  \hat{A}(\bm{q},T)^{-1} $ and solve Eq.~\eqref{GL-q} formally, 
\begin{equation}
m_{\bm{q}}^{\alpha} = - \gamma \sum_{j,\alpha'} e^{-i \bm{q} \cdot \bm{R}_j} \eta_{\alpha \alpha'} (\bm{q},T) \Pi_{\alpha', j} ,
\label{mq-chi}
\end{equation}
which with Eq.~\eqref{Pi} leads to an eigenvalue equation,
\begin{equation}
\Pi_{\alpha, j} = - \gamma \sum_{j',\alpha'} \Lambda_{jj'}^{\alpha \alpha'}(T)  \Pi_{\alpha', j'}, 
\label{Pi-eigenvalue}
\end{equation}
where we define the coupling 
\begin{equation}
\Lambda_{jj'}^{\alpha \alpha'}(T) = \int d^2 q \; e^{i \bm{q} \cdot (\bm{R}_{j} - \bm{R}_j')} \eta_{\alpha \alpha'}(\bm{q},T) \; .
\label{Lambda-eta}
\end{equation}
From App.~\ref{sec:appendixSaddle} we see that $ \eta_{\alpha \alpha'}(\bm{q}) $ is related to the spin susceptibility such that $ \Lambda_{jj'}^{\alpha \alpha'} $ has the features of an RKKY-type of coupling between impurity sites, although we deal with non-magnetic impurities. 

We solve Eq.~\eqref{Pi-eigenvalue} as an eigenvalue equation, which gives the criterion for the nucleation of magnetic order around the impurities. In particular, it allows us to determine the critical temperature $ T_N' $ (or the critical $\gamma$ for fixed temperature) together with the structure of the spin polarization pattern through the eigenvector $ \Pi_{\alpha j} $ and use of Eqs.~(\ref{m-Fourier}) and (\ref{mq-chi}). 

\begin{figure}[t!]
 \centering
\begin{minipage}[t]{0.75\columnwidth}
		\includegraphics[width=\textwidth]{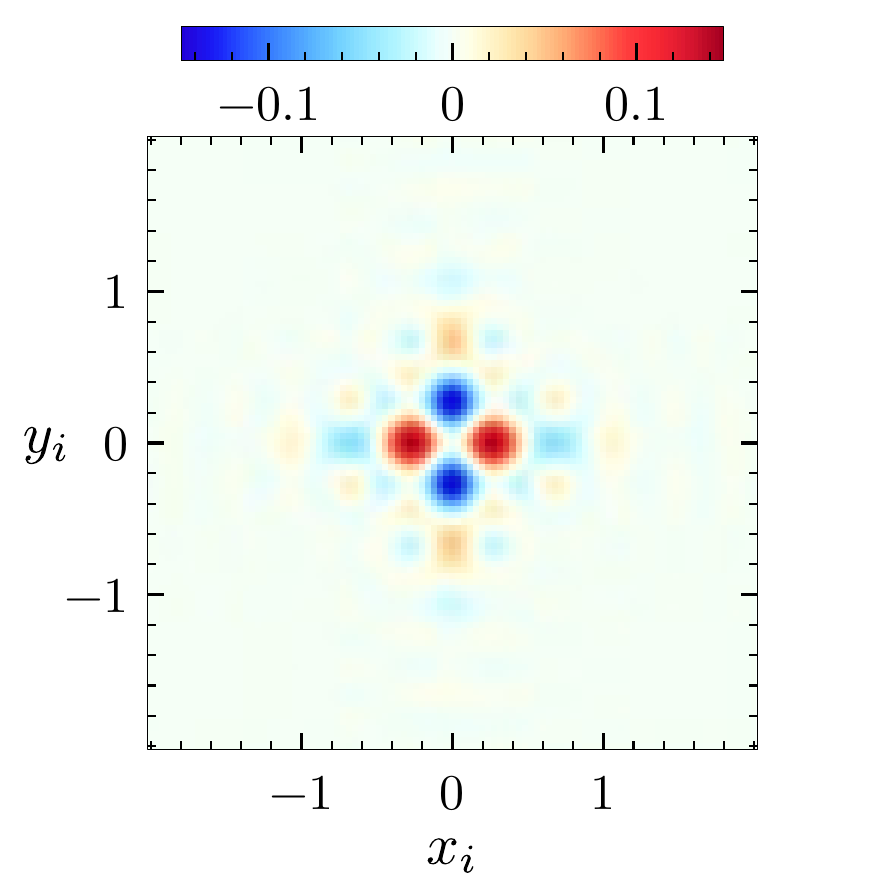}
\end{minipage}
\caption{The magnetization of both orbitals around an impurity in the center. The parameters have been chosen through comparison with the effective field theory in Sect.~\ref{sec:EFTsection}, which is based on the microscopic model. For the effective model we used $t'=\lambda = 0.1 t $, electron density $ n= 8/3 $, temperature $ T =0.1 t$ and a repulsive onsite potential $V$. The distances in real space are expressed in units of the correlation length.}
\label{fig:mf-gl-comp}
\end{figure}

Before the discussion of the multi impurity problem in the next subsection, we briefly analyze the single impurity part of the magnetic instability, which leads to the reduced equation
\begin{equation}
\sum_{\alpha'} \left[ \delta_{\alpha,\alpha'} + \gamma \Lambda_{jj}^{\alpha \alpha'}(T) \right] \Pi_{\alpha', j} =0,
\end{equation}
where 
\begin{equation}
\Lambda_{jj}^{\alpha \alpha'}(T) = \int d^2 q\; \eta_{\alpha \alpha'} (\bm{q},T) .
\end{equation}
By symmetry we find $ \Lambda(T) = \Lambda_{jj}^{xx}(T) = \Lambda_{jj}^{yy}(T) $ and $ \Lambda'(T)= \Lambda_{jj}^{xy}(T)= \Lambda_{jj}^{yx}(T) $. Taking the single impurity as the symmetry center the point group $ C_{4v} $ is preserved and we can separate the solutions according to the two representations into
\begin{align}
\Pi_{A,j} & = \frac{1}{2} ( \Pi_{x,j} + \Pi_{y,j} ), \label{xxyy} \\[3mm]
\Pi_{B,j} & = \frac{1}{2} ( \Pi_{x,j} - \Pi_{y,j} ), \label{xyyx}
\end{align}
which yields two decoupled equations,
\begin{align} 
\left[ 1 + \gamma (\Lambda_{jj}^{xx} (T)+  \Lambda_{jj}^{xy}(T)) \right] \Pi_{A,j} & = 0 , \\[3mm]
\left[ 1 + \gamma (\Lambda_{jj}^{xx} (T)-  \Lambda_{jj}^{xy}(T) ) \right] \Pi_{B,j} & = 0 .
\end{align}
The comparison with the microscopic theory shows that $ \Lambda_{jj}^{xx} >0 $, $ \Lambda_{jj}^{xy} < 0 $ and $ \Lambda_{jj}^{xx} -  \Lambda_{jj}^{xy} > 0 $. As a result, the solution of the $ B_1 $ representation would be the first to nucleate in accordance with the finding of the previous chapter. Returning to the real space magnetization using Eqs.~(\ref{m-Fourier}) and (\ref{mq-chi}), we find a cross-like pattern, as depicted in Fig.~\ref{fig:mf-gl-comp}, which is qualitatively similar to the one found in Fig.~\ref{fig:spinpol}.

\subsection{Magnetic order with several impurities}

\begin{figure}[t!]
 \centering
\begin{minipage}[t]{0.49\columnwidth}
		\includegraphics[width=\textwidth]{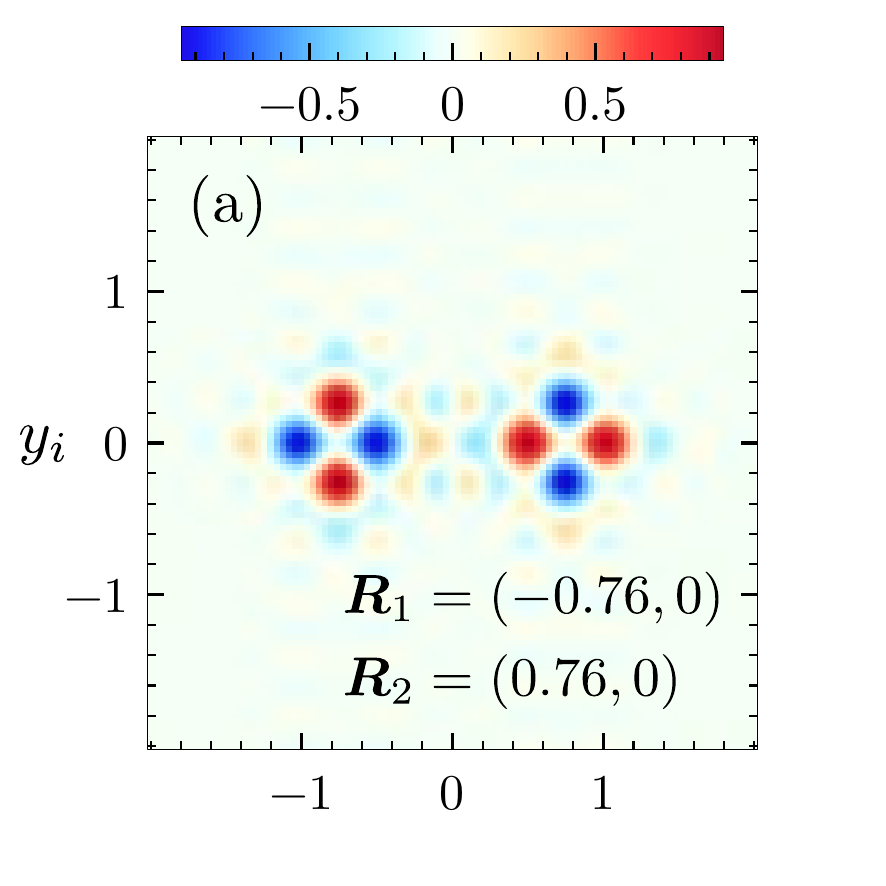}
\end{minipage}
\begin{minipage}[t]{0.49\columnwidth}
		\includegraphics[width=\textwidth]{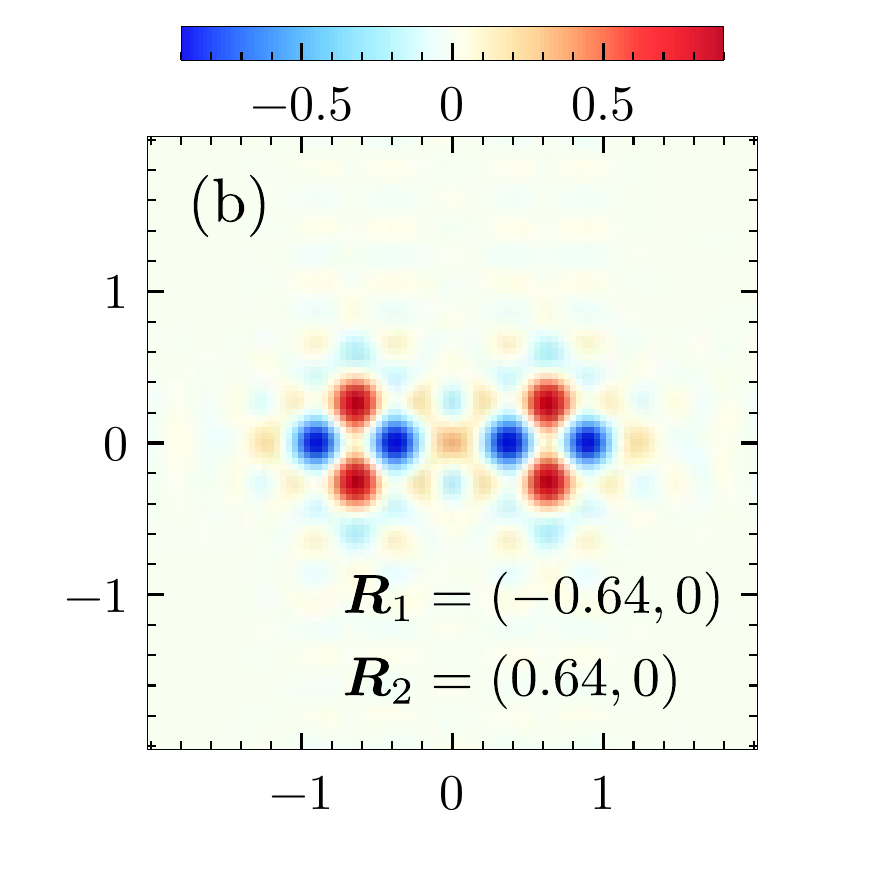}
\end{minipage}
\begin{minipage}[t]{0.49\columnwidth}
		\includegraphics[width=\textwidth]{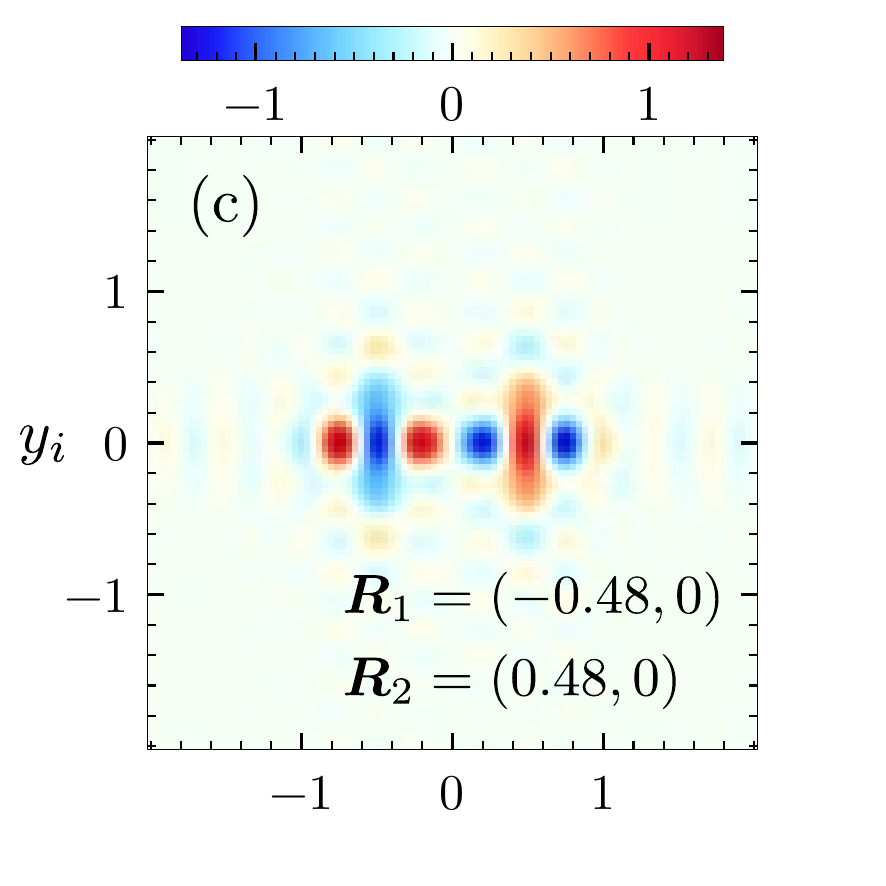}
\end{minipage}
\begin{minipage}[t]{0.49\columnwidth}
		\includegraphics[width=\textwidth]{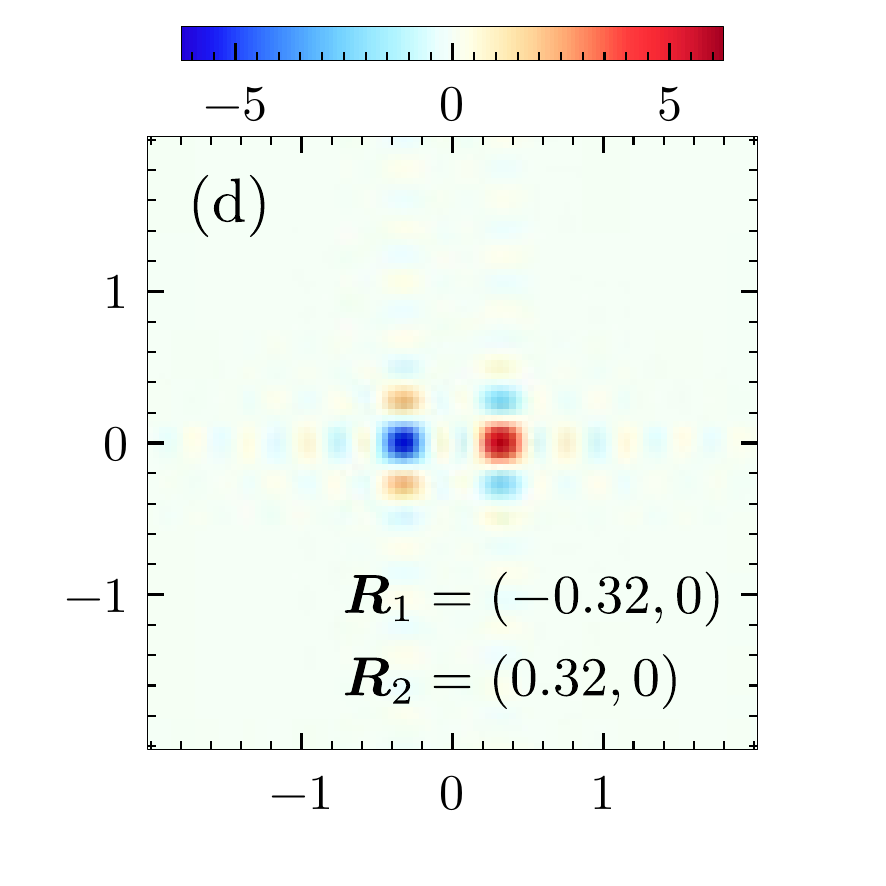}
\end{minipage}
\begin{minipage}[t]{0.49\columnwidth}
		\includegraphics[width=\textwidth]{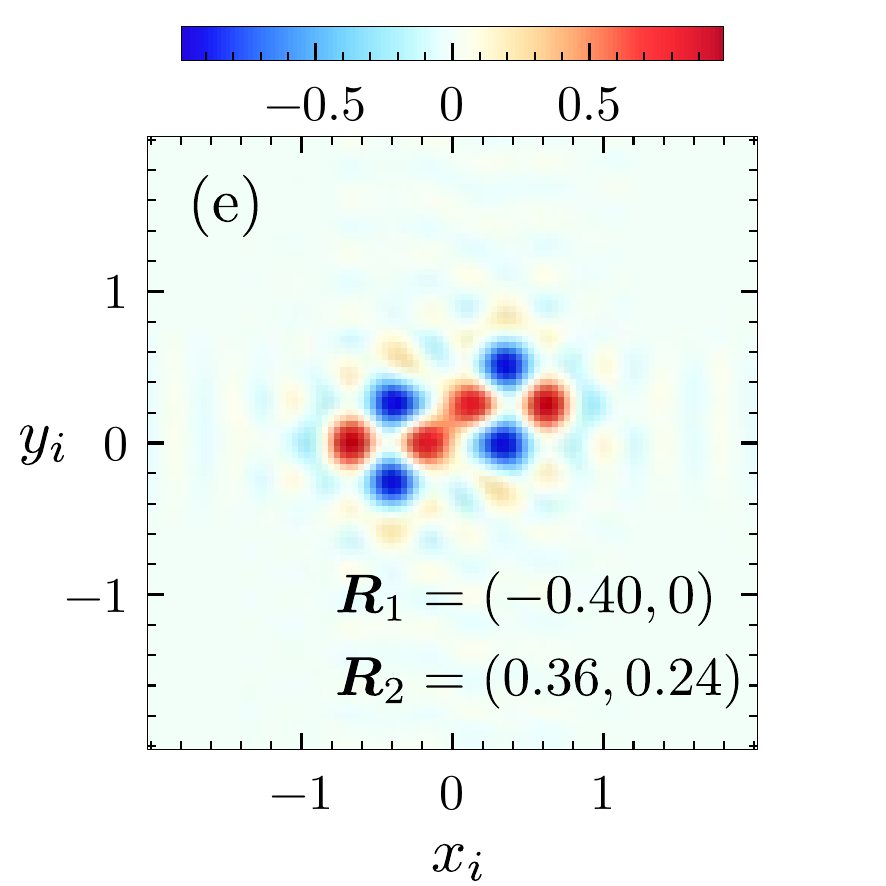}
\end{minipage}
\begin{minipage}[t]{0.49\columnwidth}
		\includegraphics[width=\textwidth]{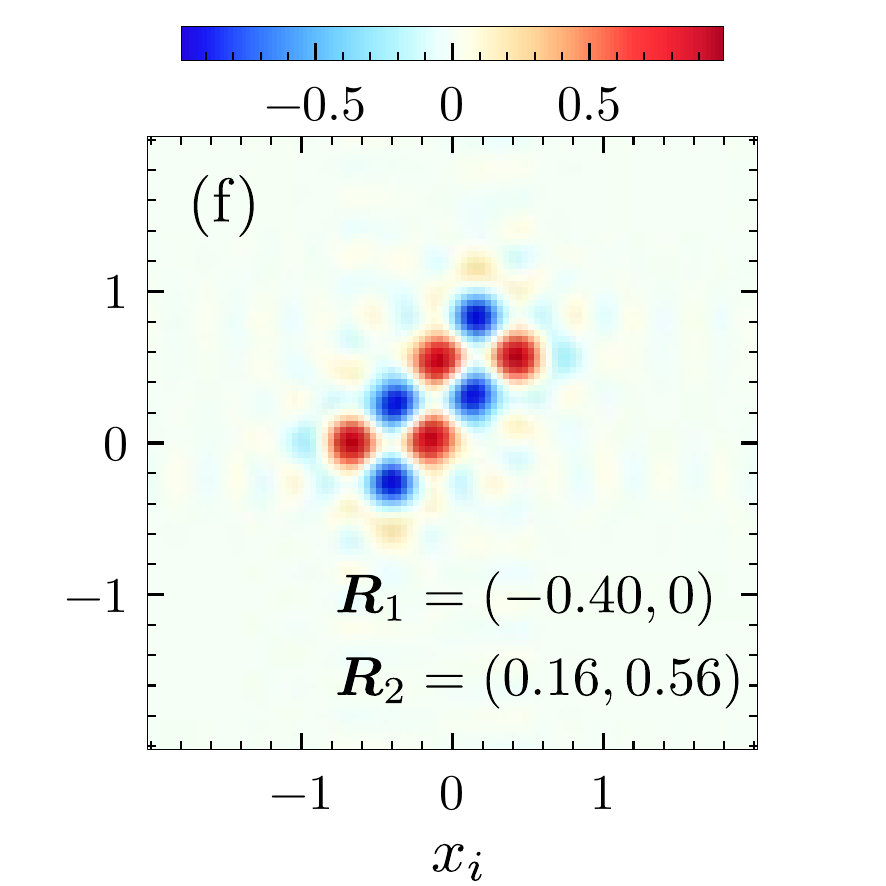}
\end{minipage}
\caption{The sum over both order parameters in real space, $m_x$ and $m_y$, derived from the GL theory for two impurities. All distances are expressed in units of the correlation length. In (a) to (d) the impurities are located on the horizontal line defined by $y_i = 0$. The magnetic moments are calculated for different distances between the impurities, which we symmetrically distributed around the origin. In (e) and (f) one impurity is rotated around the other one along a circle of radius $r = 4$. To be in line with the microscopic model, we choose for the parameters of the effective action $t'=\lambda=0.1t$, where $t$ is taken as the unit of energy, and $n=8/3$. The temperature is set to $T=0.1t$.}
\label{fig:GLshift}
\end{figure}

In order to illustrate the interplay between different impurities we consider the case of two impurities in multiple spatial configurations. In view of the solutions of the local mean field calculation in the last chapter, we use parameters yielding a correlation length of a few lattice constants at the nucleation. This means also that the inter-impurity interaction $ \Lambda_{jj'}^{\alpha \alpha'} $ decays on this length scale. Thus, we consider impurity arrangements in a comparable range of distances as depicted in Fig.~\ref{fig:GLshift} (a-f). In order to be closer to the microscopic description, we use for these simulations not the matrix $ \hat{A}(\bm{q}) $ of the Ginzburg-Landau description, but $ \hat{\tilde{A}}(\bm{q})$, which we will derive through an effective field theory in the next chapter. This has the advantage that the $ \bm{q} $-dependence is treated more adequately.  

We analyze here the nucleation of magnetic order by solving the corresponding eigenvalue equation, Eq.~(\ref{Pi-eigenvalue}), for the critical temperature, which is slightly different for each configuration. From $ \Pi_{\alpha, j} $ we obtain the real space magnetization pattern shown in Fig.~\ref{fig:GLshift}. 

It is important to note that the symmetries leading to Eqs.~(\ref{xxyy}) and (\ref{xyyx}) are lost in the presence of more than one impurity. There may be still some residual symmetries depending on the relative location of the impurities. This leads to a coupling of the two representations and a discussion in terms of $ \Pi_{x,j} $ and $ \Pi_{y,j} $ is more accessible. 

In Fig.~\ref{fig:GLshift} (a-d) we display configurations where the impurities are placed along a main axis ($x$-axis for $ y_j = 0$) at different distances. For longer distances the cross structure of each impurity is like that of a single impurity. As the impurities get closer the overlap of the magnetization pattern increases such that the beam along the $x$-axis grows in strength and the $y$-axis beam becomes comparatively weaker (note the color scales of each panel). A further feature is the change of the relative sign of the two impurity patterns, which is a result of the oscillatory RKKY-like interaction structure and connected with the wave vector $ Q_0 $. Note that we do not include the effect of electron depletion due to the impurity. Consequently, the polarization is generally not suppressed at the impurity sites. Only if the polarizations of both orbitals cancel each other, we find $m_x + m_y = 0$ at the impurity sites. Figs.~\ref{fig:GLshift} (e, f) show configurations where the two impurities are placed along the different directions. This then results in a more complex pattern, where the cross-shape however is still visible. The increased polarization between the two impurities goes in line with a slightly enhanced nucleation temperature.
The critical temperature $ T_N'$ for a single impurities is, therefore, on a mean field level a lower bound for the nucleation temperature of the many-impurity system, where the transition may be rather inhomogeneous.


\section{Effective Field theory} \label{sec:EFTsection}

In this chapter, we introduce a field theory based on the microscopic model to validate our Ginzburg-Landau approach. It will also allow us to consider further aspects of the system that have been omitted so far by including $ U(1) $ and $ SU(2) $ gauge fields, which give access to charge and spin currents in the system.

\subsection{Formulation and basic results} \label{sec:gauge_formul}

Our starting point is the partition function $Z$ expressed as a path integral~\cite{nagaosa1999quantum}, 
\begin{align}
  Z = \int \mathcal{D}C^{\dagger} \mathcal{D}C \ \text{exp} \left[ - \int_{0}^{\beta}  L(C_{\bm{k} s}^{\dagger}, C_{\bm{k} s}) d\tau  \right],
\end{align}
where $\beta = 1/T$ and the Lagrangian $L$ is derived from the Hamiltonian in Eq.~\eqref{hamil-complet}, in particular Eqs.~\eqref{H0} and  \eqref{HU}, 
\begin{equation}
 L(C_{\bm{k} s}^{\dagger}, C_{\bm{k} s}) = \sum_{\bm{k},s} C_{\bm{k}s}^{\dag} \partial_{\tau} C_{\bm{k}s} + H_0 + H_U .
\end{equation}
In a first step we perform a standard Hubbard-Stratonovich transformation decoupling the Coulomb interaction with the auxiliary field $\phi_{\bm{q}} = (\phi_{\bm{q}}^{x},\phi_{\bm{q}}^{y})^{T}$, which are conjugate fields to $ (m_{\bm{q} x}, m_{\bm{q} y}) $~\cite{nagaosa1999quantum,altlandsimons2010,mudry2014lecture}. This leads to
\begin{align}
    Z &= \int \mathcal{D}\phi \mathcal{D}C^{\dagger} \mathcal{D}C  \nonumber \\ 
    &\times \text{exp} \left[ - \int_{0}^{\beta} L'(\phi, C_{\bm{k} s}^{\dagger}, C_{\bm{k} s}) d\tau  \right] .
\end{align} 
with 
\begin{align}
     L'(\phi, C_{\bm{k} s}^{\dagger}, C_{\bm{k} s}) = &\sum_{\bm{k},\bm{q},s} C^{\dagger}_{\bm{k_{-}}s} \left( \partial_{\tau} + \hat{H}_{\bm{k},s} \right) \delta_{\bm{q},0} C_{\bm{k_{+}}s} \nonumber \\[3mm]
    &+ \sum_{\bm{k},\bm{q},s} C^{\dagger}_{\bm{k_{-}}s} \left( V_{-\bm{q}} + s \hat{\phi}_{-\bm{q}}\right) C_{\bm{k_{+}}s} \nonumber \\[3mm]
& + \sum_{\bm{q}} \sum_{\alpha, \alpha'} \phi_{\bm{q}}^{\alpha} U_{\alpha \alpha'}^{-1} \phi_{\bm{q}}^{\alpha'} .\label{eqn:HS-L}
\end{align}
The Hamiltonian matrix $ \hat{H}_{\bm{k}} $, the impurity potential $ V_{\bm{q}} $ and the interaction matrix $ \hat{U} $ are defined in Sect.~\ref{sec:modelHamil} and $ \hat{\phi}_{\bm{q}} = \textrm{diag}(\phi_{\bm{q}}^x, \phi_{\bm{q}}^y) $. 
For simplicity, we omit the charge density fields, 
$ \rho_{\bm{q}} $ [Eq.~(\ref{rho-q})], which can be absorbed in the chemical potential, and, thus, also neglect the effect of the repulsive inter-orbital coupling $ K $. We can extend the Lagrangian by introducing the electromagnetic $U(1)$ gauge field $\bm{A}$ and the spin $SU(2)$ gauge field $\bm{a}$, which couple to charge and spin currents, as will be discussed later. The complete Lagrangian is given in App.~\ref{sec:appendixSaddle} in Eq.~\eqref{eqn:fullLagr}.\par 

After integrating out the Fermionic degrees of freedom we end up with an effective action (for technical details see App.~\ref{sec:appendixSaddle}). Neglecting the gauge fields for now, we find up to second order in the auxiliary field $ \phi_{\bm{q}}^{\alpha} $,  
\begin{align}
S_{\rm eff}^{(2)} &= \sum_{\bm{q}, i \nu_l} \sum_{\alpha,\alpha'}  \phi_{q}^{\alpha} \left[ U_{\alpha \alpha'}^{-1} 
 -  \chi_{0}^{\alpha \alpha'}(q) \right] \phi_{-q}^{\alpha'} \nonumber \\
& + V \sum_{\substack{\bm{q},\bm{q}', \\ i \nu_l, i \nu_{l'}}} \sum_{j,\alpha, \alpha'} e^{i(\bm{q}-\bm{q}') \cdot \bm{R}_j } \phi_{q}^{\alpha} \Gamma_{\alpha,\alpha'} (q,q') \phi_{q'}^{\alpha'}, \label{eqn:Seff}
\end{align}
with the bare susceptibility 
\begin{align}
\chi_{0}^{\alpha \alpha'} (q) & = -\frac{1}{\beta} \sum_{\bm{k},s,i\omega_n} G_{0,s}^{\alpha \alpha'} (\bm{k},i\omega_n) \nonumber \\
\times  G_{0,s}^{\alpha' \alpha} &(\bm{k} + \bm{q}, i\omega_n + i \nu_l)  \label{eqn:bare_suscep}
\end{align}
and
\begin{align}
\Gamma_{\alpha,\alpha'} (q, q') & = \frac{1}{\beta^{3/2}}\sum_{\bm{k}, s, i \omega_n} \sum_{\tilde{\alpha} } 
G_{0,s}^{\alpha \tilde{\alpha}} (\bm{k}+ \bm{q}', i \omega_n + i \nu_{l'}) \nonumber  \\ 
\times G_{0,s}^{\tilde{\alpha} \alpha'} &( \bm{k}+ \bm{q},i \omega_n + i \nu_l) G_{0,s}^{\alpha' \alpha} ( \bm{k}, i \omega_n) . \label{eqn:gamma_a_apr}
\end{align}
Note that we use here the short notation $q = (\bm{q}, i\nu_l)$. Minimizing this action we obtain the mean field solution for $\phi_{\bm{q}}^{\alpha}$, or for $m_{\bm{q}}^{\alpha}$, using the relation,
\begin{equation}
m_{\bm{q}}^{\alpha} = - \sum_{\alpha'} U_{\alpha \alpha'}^{-1} \phi_{\bm{q}}^{\alpha'},
\end{equation}
derived in App.~\ref{sec:phi_m_rel}. From now on we only consider the static case. The variational equations are then concisely written as 
\begin{equation}
\sum_{\alpha'} \tilde{A}_{\alpha \alpha'} (\bm{q}) m_{-\bm{q}}^{\alpha'} = - V \sum_{j} e^{i \bm{q} \cdot \bm{R}_j} \tilde{\Pi}_{\alpha,j} (\bm{q}),
\label{eqn:seff-nucl-maintxt}
\end{equation}
with
\begin{equation}
\hat{\tilde{A}}(\bm{q}) = \hat{\tilde{\chi}}(\bm{q})^{-1} = \left[1 - \hat{U} \hat{\chi}_0(\bm{q}) \right] \hat{U} \label{eqn:gauge_matrixA}
\end{equation}
and
\begin{align}
\tilde{\Pi}_{\alpha,j} (\bm{q}) = \sum_{\bm{q}'} \sum_{\alpha',\alpha'',\alpha'''} &e^{-i\bm{q}'\cdot \bm{R}_j} U_{\alpha \alpha'} \Gamma_{\alpha' \alpha''}(\bm{q},\bm{q}')\nonumber \\ 
&\times  U_{\alpha'' \alpha'''} m_{-\bm{q}'}^{\alpha'''}.
\end{align}
Obviously, the renormalized coupling $ \hat{\tilde{A}}(\bm{q}) $ incorporates the random phase approximation for the susceptibility [see also Eq.~(\ref{m-rpa})].  
Note that Eq.~\eqref{eqn:seff-nucl-maintxt} corresponds directly to the variational equations obtained within the Ginzburg-Landau theory [Eq.~\eqref{GL-q}]. However, we should be aware that $ \hat{A}(\bm{q}) $ is only a qualitative approximation of $ \hat{\tilde{A}}(\bm{q}) $, since it is a small-$\bm{q}$ expansion, but attempts to mimic the behavior around $ Q_0 $. 
After a comparison we find for the coefficients of the gradient terms of the Ginzburg-Landau free energy [Eq.~\eqref{GL-FE}] that $ K_1 \sim K_3 \gg K_2 $. The essentially vanishing $ K_2 $ ensures the clear cross-structure. 

\subsection{Correlation functions connecting gauge fields and potentials}\label{correl-sect}
\begin{figure}[t!]
 \centering
	\includegraphics[width=0.90\columnwidth]{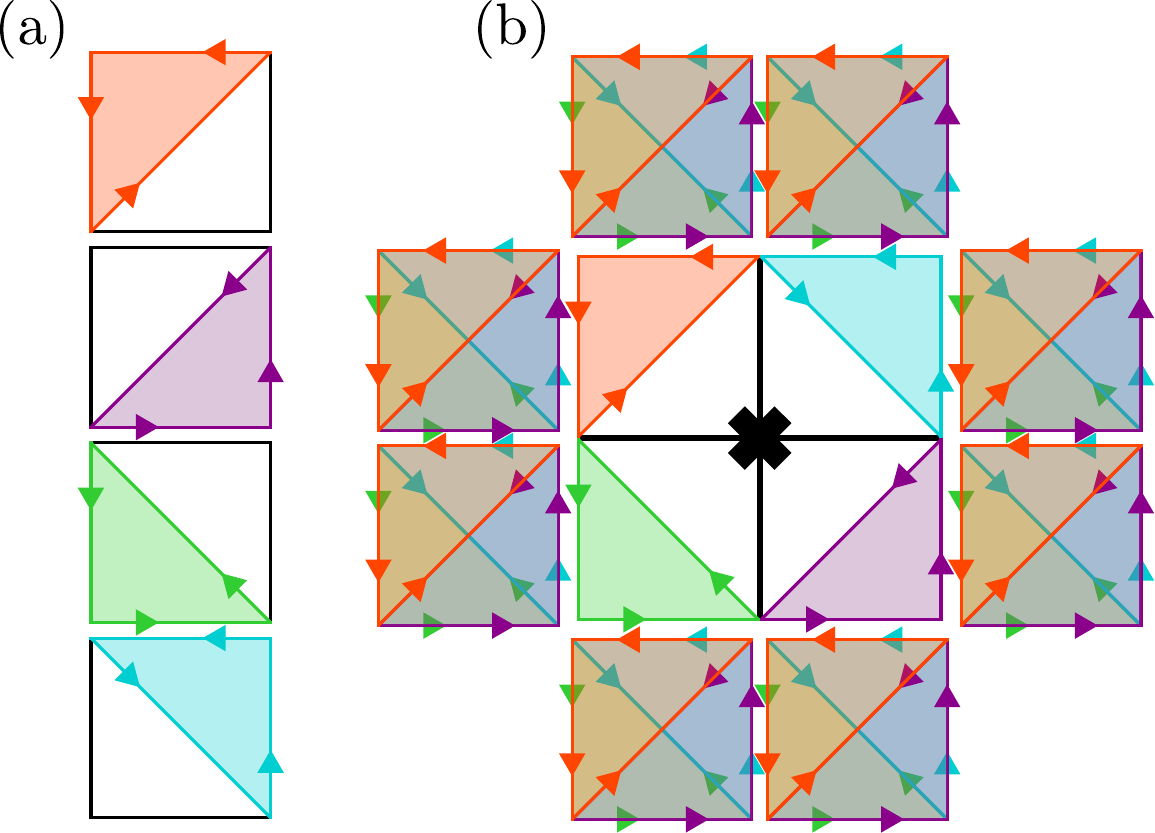}
	\caption{
		Schematic spin current pattern around a single impurity neglecting the local charge density modulation. In (a) we plot all possible triangle paths for a single electron with spin $s$ and in (b) the resulting spin current pattern around the impurity (black cross).
	}
	\label{fig:cartoon}
\end{figure}

The Hamiltonian  $ {\cal H}_b + {\cal H}_{SO} $ in Eqs.~(\ref{hb}) and (\ref{hso}) incorporates a peculiar feature. The topology of the tight-binding matrix contains a hopping 
by one lattice constant along the $ x $($y$)-direction for the $x$($y$)-orbital. Together with the diagonal inter-orbital hopping in the square plaquette a triangle path is formed. It can be closed in the plaquette corner by spin-orbit coupling connecting the two orbitals. Since spin-orbit coupling introduces a phase $ s \pi/2 $ depending on the spin $s$, the triangles host a circulating spin current  (see Fig.~\ref{fig:cartoon}). Combining all four triangles of a plaquette leads to a net circular current on the edges, while it cancels on the diagonals. In the homogeneous system the circular currents of neighboring plaquettes compensate each other. Interestingly, an impurity potential on a site interrupting four closed triangular paths leads to a net circular spin current around the impurity, as illustrated in Fig.~\ref{fig:cartoon}. 

This effect, which is due to the spin-orbit coupling and the specific form of the  inter-orbital hybridization, can be incorporated into our effective field theory by means of gauge fields. In Eq.~\eqref{M-expansion} we have included the $U(1)$ gauge field $\bm{A}$ and the $SU(2)$ gauge field $\bm{a}$, where for the latter we restricted ourselves to the $z$-axis component of the spin. 
The expansion to lowest order leads to new terms, which couple the gauge fields to the potentials for the charge and spin density. 
The extended effective action takes the form 
\begin{align}
S_{\rm eff} = S_{\rm eff}^{(2)} + S_{\rm eff}^{\rm g},
\end{align}
with
\begin{align}
S_{\rm eff}^{\rm g} &= \sum_{\bm{q}, i \nu_l} \sum_{\alpha,\mu} \left[ A_{q}^{\mu} C_{\alpha}^{\mu}(q) \phi_{-q}^{\alpha} 
 + a_{q}^{\mu} C_{\alpha}^{\mu}(q) V_{-\bm{q}} \right] \nonumber \\
& + \sum_{\bm{q}, i \nu_l} \sum_{\mu, \mu'} \left[ A_{q}^{\mu} \tilde{C}^{\mu \mu'}(q) A_{-q}^{\mu'} + a_{q}^{\mu} \tilde{C}^{\mu \mu'}(q) a_{-q}^{\mu'}\right], \label{eqn:Seff_gauge}
\end{align}
where we introduce the correlation functions
\begin{align}
C_{\alpha}^{\mu}(q) &= \frac{1}{\beta} \sum_{\bm{k},s,i\omega_n} \sum_{\alpha', \tilde{\alpha}} G_{0,s}^{\alpha \tilde{\alpha}} (\bm{k}_{-},i\omega_n) s \left(\hat{I}_{\bm{k}}^{\mu}\right)^{\tilde{\alpha} \alpha'} \nonumber \\ 
&\times G_{0,s}^{\alpha' \alpha} (\bm{k}_{+}, i\omega_n + i \nu_l), 
\end{align}
and 
\begin{align}
\tilde{C}^{\mu \mu'}(q) &= \frac{1}{\beta} \sum_{\bm{k},s,i\omega_n} \sum_{\alpha, \alpha', \tilde{\alpha}, \tilde{\alpha}'} G_{0,s}^{\alpha \alpha'} (\bm{k}_{-},i\omega_n)  \left(\hat{I}_{\bm{k}}^{\mu'}\right)^{\alpha' \tilde{\alpha}} \nonumber \\ 
&\times G_{0,s}^{\tilde{\alpha} \tilde{\alpha}'} (\bm{k}_{+}, i\omega_n + i \nu_l) \left(\hat{I}_{\bm{k}}^{\mu}\right)^{\tilde{\alpha}' \alpha}. 
\end{align}
We have introduced here  the current matrix in $\bm{k}$-space, defined through the relation
\begin{align}
    \hat{\bm{I}}_{\bm{k}} = \frac{1}{\hbar} \bm{\nabla}_{\bm{k}} \hat{H}_{\bm{k},s}. 
\end{align}
Setting $\hbar = 1$ we find for the $x$- and $ y$-component
\begin{align}
    \hat{I}_{\bm{k}}^{x} = \begin{pmatrix}
    	V_{\bm{k}}^x & W_{\bm{k}}^x \\[2mm]
    	W_{\bm{k}}^x & 0
    \end{pmatrix}, \quad 
    \hat{I}_{\bm{k}}^{y} = \begin{pmatrix}
    	0 & W_{\bm{k}}^y \\[2mm]
    	W_{\bm{k}}^y & V_{\bm{k}}^y
    \end{pmatrix}, \label{eqn:current_matrix}
\end{align}
with $V_{\bm{k}}^x = 2t \sin k_x$, $V_{\bm{k}}^y = 2t \sin k_y$, $W_{\bm{k}}^x = 4t' \cos k_x \sin k_y$ and $W_{\bm{k}}^y = 4t' \cos k_y \sin k_x$. These current matrices involve both the nearest and next-nearest neighbor contributions and do not depend on the spin $ s $. 

Note that in Eq.~\eqref{eqn:Seff_gauge} we have already omitted those terms which vanish entirely after summing over ($\bm{k}$, $n$, $s$). The finite terms are summarized in Table~\ref{tab:Seff_gaugeterms}. 

\begin{table}[h!]
	\begin{tabular}{| c | c | c |}
	 \hline 
	  $S_{\rm eff}^{\rm g}$ & CF & \rule{0pt}{4mm} \\ [1.5mm]
 	 \hline 
 	 \begin{tabular}{@{}c@{}} \rule{0.mm}{0pt} $A_{q}^{\mu} C_{\alpha}^{\mu}(q) \phi_{-q}^{\alpha}$ \rule{0pt}{2mm}  \\[4mm] \rule{0.mm}{0pt} $a_{q}^{\mu} C_{\alpha}^{\mu}(q) V_{-\bm{q}} $ \end{tabular} & $~C_{\alpha}^{\mu}(q)$ \rule{0pt}{10mm} & \begin{tabular}{@{}c@{}} \rule{0.3mm}{0pt} $\text{CC}-\text{SD} $ \rule{0pt}{2mm}  \\[4mm] $\text{SC}-\text{CD}$\end{tabular}  \\ [6.0mm]
 	  \hline \hline
 	   \begin{tabular}{@{}c@{}} \rule{0.mm}{0pt} $A_{q}^{\mu} \tilde{C}^{\mu \mu'}(q) A_{-q}^{\mu'}$ \rule{0pt}{2mm}  \\[4mm] \rule{0.mm}{0pt} $a_{q}^{\mu} \tilde{C}^{\mu \mu'}(q) a_{-q}^{\mu'}$ \end{tabular}
     & $~\tilde{C}^{\mu \mu'}(q)$ \rule{0pt}{10mm} & \begin{tabular}{@{}c@{}} $\text{CC} - \text{CC}$ \rule{0pt}{2mm}  \\[4mm] $\text{SC} - \text{SC}$\end{tabular}  \\ [7mm]
     \hline
	\end{tabular}
	  \caption{
		The lowest order coupling terms of $S_{\rm eff}^{\rm g}$, which appear by including the $U(1)$ gauge field $\bm{A}$ and the $SU(2)$ gauge field $\bm{a}$, and their respective correlation functions (CF). The origin of each term is given in the last column (CD: charge density, SD: spin density, CC: charge current, SC: spin cur- rent).}
		\label{tab:Seff_gaugeterms}
\end{table}\par 
We will work in the following with the density-current part and list the current-current contribution just for completeness. The correlation function $ C_{\alpha}^{\mu} (q) $ has the form
\begin{align}
C^x_{y} (\bm{q}) = - \frac{1}{\beta N^2} \sum_{k} & \frac{i \lambda}{{\cal D}_{k_+} {\cal D}_{k_-}} \left[ (\gamma_{\bm{k}_+} - \gamma_{\bm{k}_-}) V_{\bm{k}}^x \right. \nonumber \\ & \left. - (\xi_{y, \bm{k}_+} - \xi_{y, \bm{k}_-}) W_{\bm{k}}^x \right]
\end{align}
and
\begin{align}
C^y_{x} (\bm{q}) =  \frac{1}{\beta N^2} \sum_{k} & \frac{i \lambda}{{\cal D}_{k_+} {\cal D}_{k_-}} \left[ (\gamma_{\bm{k}_+} - \gamma_{\bm{k}_-}) V_{\bm{k}}^y \right. \nonumber \\ & \left. - (\xi_{x, \bm{k}_+} - \xi_{x, \bm{k}_-}) W_{\bm{k}}^y \right],
\end{align}
while $ C^x_x (\bm{q}) = C^y_y(\bm{q}) = 0 $. It is worth noting that these correlation functions include in the denominator the combination of hoppings along the main axes ($ \xi_{x/y,\bm{k}} \to t $) and the plaquette diagonals ($ \gamma_{\bm{k}} \to t' $), as well as the spin-orbit coupling ($ \lambda $), which constitute exactly the ingredients used above for the circular spin currents in Fig.~\ref{fig:cartoon}. 

A straightforward symmetry analysis of the expressions for $ C^x_{y} (\bm{q}) $ and $ C^y_{x} (\bm{q}) $
shows that
\begin{equation}
\left( \begin{array}{c} C_y^x(\bm{q}) \\ [2mm] C_x^y(\bm{q}) \end{array} \right) = \left( \begin{array}{c} - i q_y \Upsilon(q_x,q_y) \\[2mm]  i q_x \Upsilon(q_y,q_x) \end{array} \right),
\end{equation}
where the function $ \Upsilon(q_x,q_y) $ is even in $ \bm{q} $ and  $ \Upsilon(-q_x,q_y) = \Upsilon(q_x,q_y) $, $  \Upsilon(0,0) = 0 $ and generally $  \Upsilon(q_x,q_y) \neq  \Upsilon(q_y,q_x) $. We will use this property to consider features such as the spin currents around an impurity as anticipated above, but also the connection between spin density polarization and charge currents, which are suggested by our gauge field terms. 

\subsubsection{Impurity induced spin currents}

\begin{figure}[h!]
 \centering
	\includegraphics[width=0.75\columnwidth]{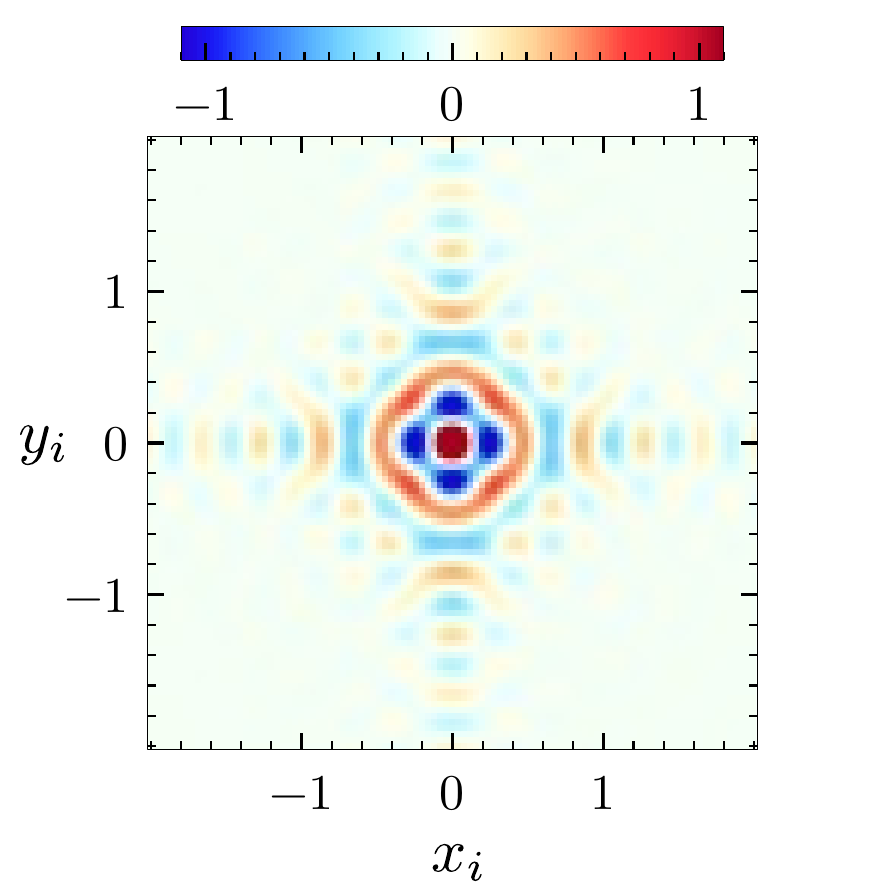}
	\caption{The vorticity $\Omega^s(\bm{r})$ of the impurity induced spin currents. The distances are given in units of the correlation length. In our calculation we used for the parameters of the effective action $t'=\lambda=T=0.1t$ with $n=8/3$. 
	}
	\label{fig:vort_SC}
\end{figure}

We first consider the spin current pattern due to a point like impurity potential. In this case the potential is structureless, $ V_{\bm{q}} = V $, and the spin current is simply obtained as
\begin{equation}
\left( j_x^s(\bm{r}) ,  j_y^s(\bm{r} \right) = V \int d^2 q \left(C_y^x (\bm{q}) , C_x^y (\bm{q}) \right) e^{-i \bm{q} \cdot \bm{r}}.
\end{equation}
It is actually easier to consider the vorticity of this current rather than the current pattern itself, i.e. $ \Omega^s (\bm{r}) = [\bm{\nabla} \times \bm{j}^s(\bm{r}) ]_z $,
\begin{align}
&\Omega^s (\bm{r}) = -i V \int d^2 q \; \left[ q_x C_x^y (\bm{q}) - q_y  C_y^x (\bm{q})) \right] e^{-i \bm{q} \cdot \bm{r}} , \nonumber \\[3mm]
&\hspace{0.5cm}= V \int d^2 q \left[ q_x^2 \Upsilon(q_y,q_x) + q_y^2 \Upsilon(q_x,q_y) \right]e^{-i \bm{q} \cdot \bm{r}} . \label{spin-vorticity}
\end{align}
It is obvious that $ q_x^2 \Upsilon(q_y,q_x) + q_y^2 \Upsilon(q_x,q_y)  $ has the full symmetry of the point group $ C_{4v} $. The vorticity $ \Omega^s (\bm{r}) $ 
as a function of $ \bm{r} $ has thus the same symmetry, which belongs to the irreducible representation $ A_1 $, as Fig.~\ref{fig:vort_SC} displays. 
This means that we find a circular spin current around the impurity site analogous to the schematic picture shown in Fig.~\ref{fig:cartoon}. However, the correlation function has the property that it peaks for $ q_x \approx \pm Q_0 $ and $ q_y \approx \pm Q_0  $, which yields an oscillation with distance such that the spin current reverses sign several times for increasing $ r $ until it fades away. It is also obvious from Eq.~(\ref{spin-vorticity})
that the net (integrated) vorticity vanishes. 

\par

\begin{figure}[t!]
 \centering
	\includegraphics[width=0.70\columnwidth]{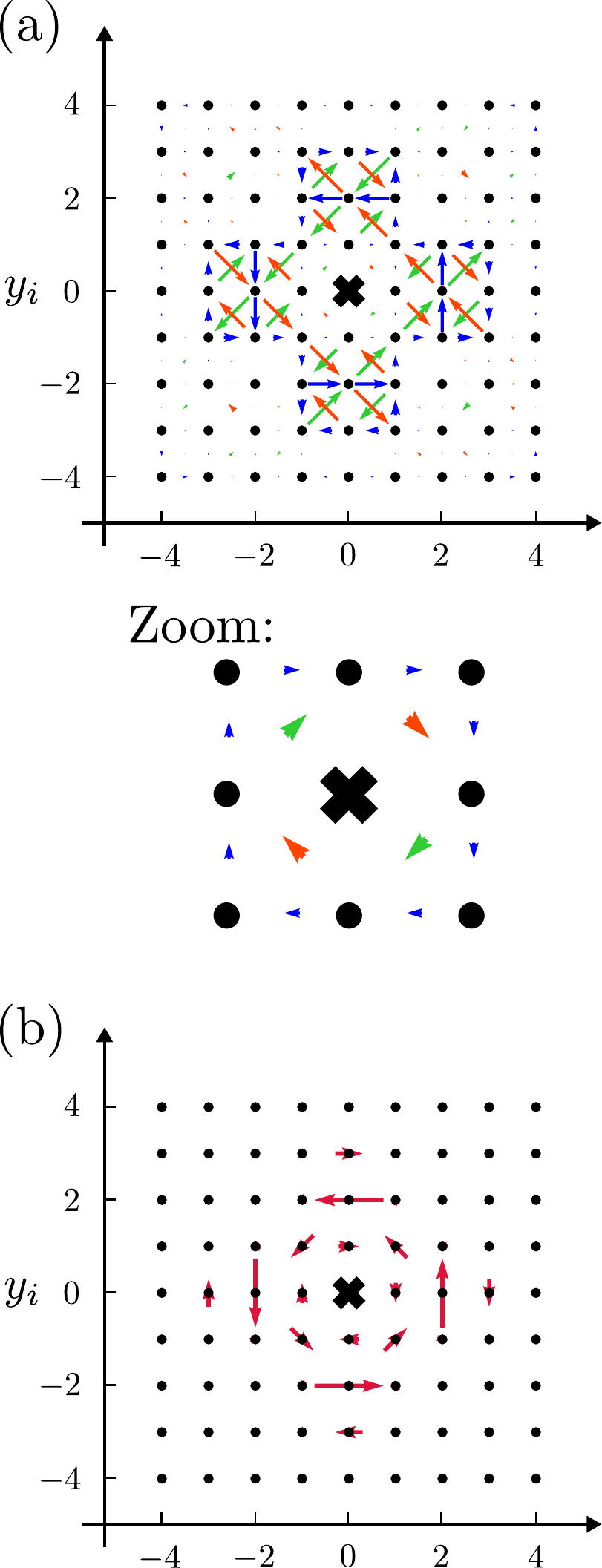}
                
\caption{The spin currents calculated with the microscopic lattice model in the absence of interactions ($U=K=J=0$). The black cross in the middle marks the impurity site and the black dots the lattice sites. In (a) we depict the currents across each bond and in (b) the net current at each lattice site. The model parameters are the same as in Sec.~\ref{sec:modelHamil}.
}
\label{fig:bondcurrents}
\end{figure}

Furthermore, it is possible to compare these findings for our square lattice model numerically, neglecting interactions. For this purposed we define the current operators for nearest neighbor bonds as
\begin{align}
    \widehat{\bm{J}}_{\bm{r} \alpha s}^{nn} = i t \left( \hat{\bm{x}} c_{\bm{r} + \hat{\bm{x}}\alpha s}^{\dagger} c_{\bm{r} \alpha s} + \hat{\bm{y}} c_{\bm{r}+\hat{\bm{y}} \alpha s}^{\dagger} c_{\bm{r} \alpha s} - h.c. \right) 
\end{align}
and for next-nearest neighbor bonds analogously as
\begin{align}
 \widehat{\bm{J}}_{\bm{r} \alpha s}^{nnn} = \frac{it'}{\sqrt{2}} & \left\{ (\hat{\bm{x}} + \hat{\bm{y}}) c_{\bm{r} + \hat{\bm{x}}+\hat{\bm{y}}  \alpha s}^{\dagger} c_{\bm{r} \alpha s} \right. \nonumber \\[2mm]
 &  \left. +  (\hat{\bm{x}} - \hat{\bm{y}}) c_{\bm{r} + \hat{\bm{x}}-\hat{\bm{y}}  \alpha s}^{\dagger} c_{\bm{r} \alpha s} - h.c. \right\} . 
 \end{align}
This allows us to obtain charge (spin) currents by adding (subtracting) the bond currents of different spins. Using this formulation we
determine the bond spin currents around a single impurity and obtain the pattern shown in Fig.~\ref{fig:bondcurrents}. Close to the impurity the current pattern agrees qualitatively well with the picture of Fig.~\ref{fig:cartoon}, as shown in the enlarged figure (b). However, further away the bond current pattern becomes more difficult to analyze. A better view is obtained by considering the spin current density passing through the sites which is obtained by simply averaging over all bonds adjacent to a site. In this way the circular character of the current is well visible, including its oscillations in size and orientation. Within our numerical calculations we have tested that the sign change of either $ t' $ (inter-orbital hopping) or the spin orbit coupling $ \lambda $ lead to a reversal of the current pattern.

\subsubsection{Charge currents induced by magnetization pattern}
\begin{figure}[h!]
 \centering
	\includegraphics[width=0.75\columnwidth]{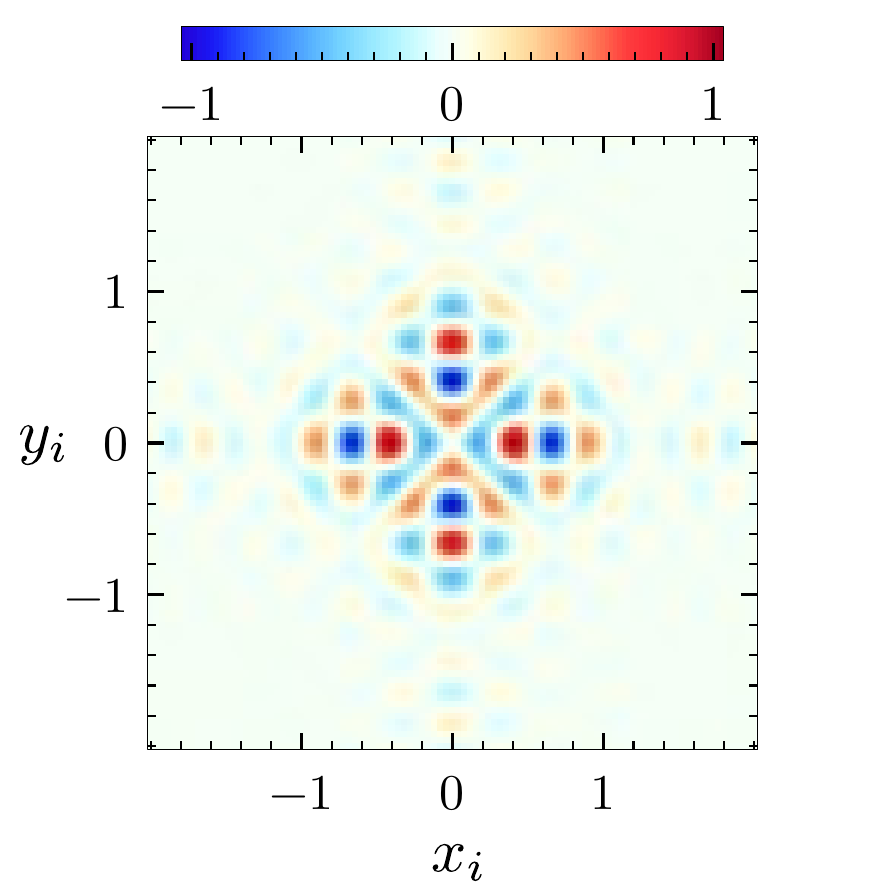}
	\caption{The vorticity $\Omega^c(\bm{r})$ of the charge currents induced by the quadrupolar magnetization pattern around each impurity. The distances are given in units of the correlation length and the parameters of the effective action are again $t'=\lambda=T=0.1t$ with $n=8/3$.
	}
	\label{fig:vort_CC}
\end{figure}

Apart from the impurity induced spin currents, we also see that the spin polarization induces charge currents, which are derived from the variation of $ S_{\rm eff}^{\rm g} $ with respect to the vector potential $ \bm{A} $,
\begin{align}
&\left( j_x^c(\bm{r}) ,  j_y^c(\bm{r} \right) \nonumber \\[3mm]
&\hspace{1cm}= \int d^2 q \left(C_y^x (\bm{q}) \phi^y_{-\bm{q}}, C_x^y (\bm{q}) \phi^x_{-\bm{q}}\right) e^{-i \bm{q} \cdot \bm{r}} .
\label{j-phi}
\end{align}
It is again illustrative to look at the vorticity, 
\begin{align}
&\Omega^c(\bm{r})  = -i \int d^2 q \left[ q_x C_x^y(\bm{q}) \phi^x_{-\bm{q}} - q_y C_y^x \phi^y_{-\bm{q}} \right] e^{-i \bm{q} \cdot \bm{r}}, \nonumber \\[3mm]
 & = \int d^2 q \left[ q_x^2 \Upsilon(q_y,q_x)  \phi^x_{-\bm{q}} + q_y^2 \Upsilon(q_x,q_y)  \phi^y_{-\bm{q}} \right]  e^{-i \bm{q} \cdot \bm{r}} . 
\end{align}                             
We consider now the pattern around a single impurity. For $ \phi_{\bm{q}}^{\alpha} $, we have then the relation
\begin{equation}
\phi_{q_x,q_y}^{x} = - \phi_{q_y,q_x}^y,
\end{equation}
where both are even functions of $ \bm{q} $. In this way, it is obvious that the expression in the integrand has the structure of a basis function of $ B_1 $ in $ C_{4v} $, i.e. like $ x^2 - y^2 $. This is reflected by the vorticity pattern illustrated in Fig.~\ref{fig:vort_CC}, which clearly exhibits the same $ B_1 $ symmetry. The charge currents generate the same magnetic field pattern as the spin magnetization. 

\subsubsection{Spin polarization induced by external charge currents}

\begin{figure}[t!]
 \centering
	\includegraphics[width=0.65\columnwidth]{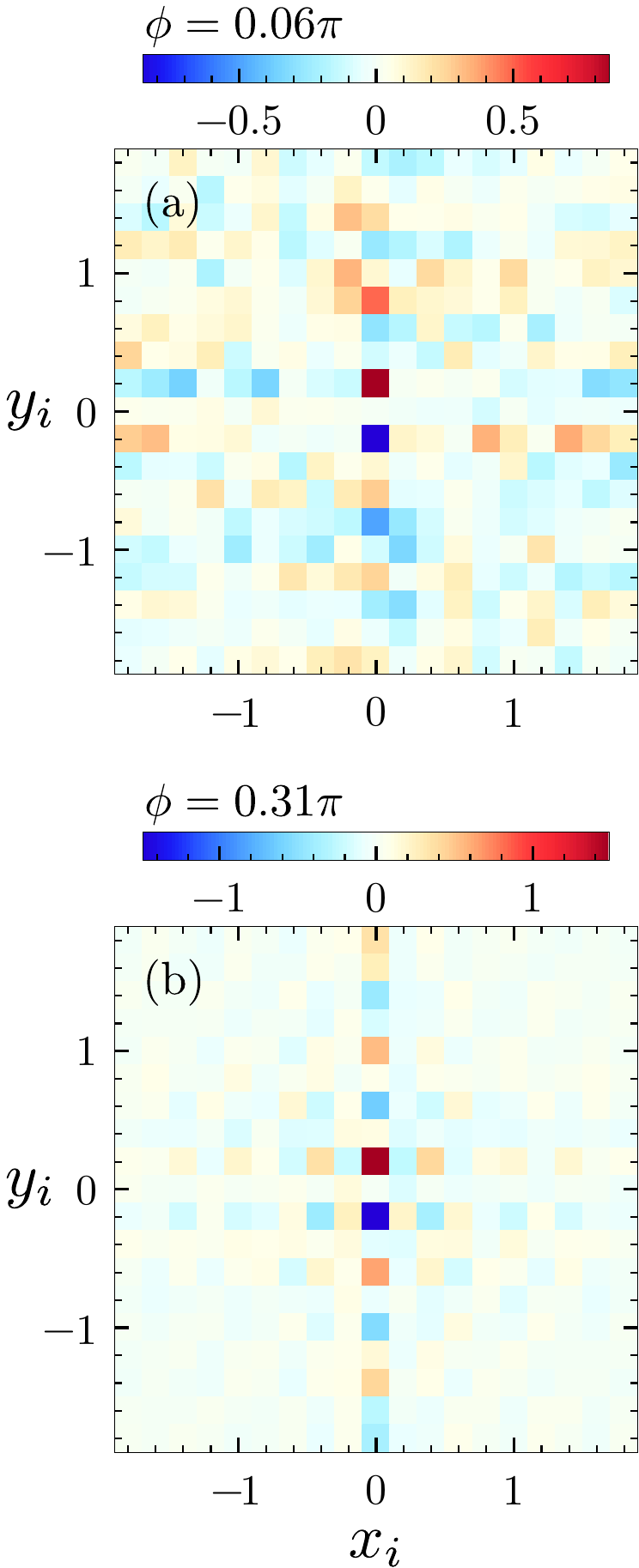}
                
\caption{The charge current induced spin polarization around the impurity in the non-interacting case. The current was applied along the $x$-axis and we calculated the polarization for two different phases $\phi = \text{Arg}(1+i A_x)$. We used the microscopic model parameters given in Sec.~\ref{sec:modelHamil}. 
}
\label{fig:currentind}
\end{figure}

An interesting effect  of the interplay between current and spin polarization is found when we consider an external current running through the system, which induces a spin polarization above the nucleation temperature of the magnetization. To illustrate this feature we use again the expression for the charge currents given in Eq.~(\ref{j-phi}). Instead of looking at the reaction of the system to a charge current, we determine the current density for a specific spin polarization, which extends from the impurity (at the origin) along the $y$-axis and has opposite signs for positive and negative $y$. Thus, we assume that 
\begin{equation}
\phi^y_{-\bm{q}} = q_y f(q_y),
\end{equation}
which involves strong localization along the $x$-direction. Inserting $ \phi^y_{-\bm{q}} $ and taking into account that the current is source and drain free, i.e. $ \bm{\nabla} \cdot \bm{j}^c(\bm{r}) = 0 $, requires that also $ \phi^x_{-\bm{q}} $ is finite. This leads to the current densities
\begin{align}
j_x^c(\bm{r}) &= \int d^2 q \; q_y^2 \Upsilon(q_x,q_y) f(q_y) e^{-i \bm{q} \cdot \bm{r}}, \\[3mm]
j_y^c(\bm{r}) &= -\int d^2 q \; q_x q_y \Upsilon(q_x,q_y) f(q_y) e^{-i \bm{q} \cdot \bm{r}} .
\end{align}
The symmetry of the current density corresponds to the pattern of a flow passing an obstacle (impurity) with
\begin{align}
j_x^c (x,y) & = j_x^c (x,-y)=j_x^c (-x,y) = j_x^c (-x,-y) , \\[3mm]
j_y^c (x,y) & = - j_y^c (x,-y)= - j_y^c (-x,y) = j_y^c (-x,-y).
\end{align}
We may also invert the point of view. This means that a uniform current parallel to a main axis in our system would induce a magnetization pattern around an impurity, which is perpendicular to the current and antisymmetric under reflection at the current axis. 

To test this we go back to our mean field discussion of the microscopic model and consider a square shaped system with one impurity at the center. 
We induce a current by means of a uniform vector potential along the $x$-direction (Peierl's phase $ \phi $ on nearest-neighbor hopping matrix elements), which then adapts to the situation around the impurity site. Note that for technical reasons we average over the twisted boundary conditions in order to reduce finite-size effects. 

In accordance with our previous discussion, a spin polarization pattern emerges around the impurity, which extends as a narrow beam along the $y$-direction starting at the impurity and has opposite signs on the two sides of the impurity, as depicted in Fig.~\ref{fig:currentind}. Evidently, the impurity acts again as a nucleation center facilitating the appearance of a finite spin polarization. 
The results shown are for two different currents, which correspond to the Peierl's phases $ \phi =0.06 \pi $ and $ 0.31 \pi $. Analyzing the model parameter dependence, we observe that the magnitude of the pattern depends not only on the strength of the current, but also strongly on $ t' $ and $ \lambda $ such that the pattern vanishes when one or both parameters are zero. This affects also the distribution of the spin polarization on the two orbitals.

\section{Conclusion} \label{sec:conclusion}

The goal of our study has been to find the origin and structure of the magnetic order induced by non-magnetic impurities in Sr$_2$Ru$_{1-x}$Ti$_x$O$_4$. Surprisingly, already a few percent of Ti-ions replacing Ru-ions lead to the magnetic instability \cite{Minakata:2001,Braden:2002Ti,Iida:2012}, which can be described well within a relatively simple model of two bands derived from the $ 4d $-$t_{2g} $ orbitals, $ d_{yz} $ and $ d_{zx} $, of the Ru-ions. 
The pronounced nesting feature of the two bands explains the spatial modulation of the magnetization with wave vectors $ |\bm{Q_0}| \approx 2 \pi/3 $. Moreover, spin-orbit coupling is an additional ingredient, which favors the spin polarization to be along the $z$-axis, as discussed in Ref.~\cite{Ng:2000}. Both features agree very well with experimental findings \cite{Braden:2002Ti}.

Because a simple impurity averaged approach is not capable of capturing the impurity induced magnetic order, we concluded that a more appropriate technique is necessary. Restricting ourselves to the two relevant bands, we could show based on a position-dependent mean field theory that a repulsive impurity potential acts as a nucleation center for the spin polarization. Around a single Ti-impurity the magnetization pattern takes a cross-like structure with beams along the crystalline main axis, each of which is dominated by one of the two involved orbitals. The orientation of the magnetization of the two beam directions is opposite such that no net magnetic moment arises. For these properties the inter-orbital hybridization as well as the spin-orbit coupling are essential. The mean field calculation also suggests that the nucleation of magnetic order occurs at a temperature higher than the transition to the incipient, uniform, incommensurate magnetic order. This would be consistent with the notion of impurity induced order.  

The interplay between different impurities is a complex problem. In order to analyze it we introduced a phenomenological approach based on a Ginzburg-Landau free energy functional with a magnetic, two component order parameter, $ m_x $ and $ m_y $, for the $z$-axis oriented magnetization of the two orbitals, $ d_{xz} $ and $ d_{yz} $, respectively. It is rather straightforward to formulate a free energy functional reproducing the single impurity result of the space-dependent mean field theory. Considering two impurities, we find that their relative position and orientation are highly important for the interaction. While the beams along the crystal axis and their overlap play a crucial role, also the matching of the nesting wave vector with the separation of the impurities influences the emerging pattern. The analysis of the interaction incorporates features reminiscent of the RKKY-type coupling among localized magnetic moments in a metal, despite the fact that Ti enters as a nominally non-magnetic impurity. We show that one can map the Ginzburg-Landau model to a model, where each impurity has two magnetic moments along the $z$-axis, one for each orbital. 

The Ginzburg-Landau functional can be derived through a field theory based on the microscopic model, which yields a more adequate spatial structure of the magnetization. 
It confirms the basic structure and adds details about the phenomenological parameters. In particular, it is obvious that the repulsive potential of the Ti-impurity yields an ''attractive'' nucleation center for the magnetic moments. 

The field theory allows us to uncover a few further aspects of the system, which were not noticed in the mean field theory and the Ginzburg-Landau treatment. After adding the $ U(1) $ and $ SU(2) $ gauge fields to the field theory we examined the effects of charge and spin currents. The particular topology of orbital hybridization and spin-orbit coupling leads to circular spin currents around the impurities, as any potential changing the electron density leads to spin currents. This kind of spin currents has been previously noticed at surfaces \cite{Imai-2012}. 
At the same time, we find that a non-vanishing magnetization yields charge currents. We have investigated these for 
a single impurity, where the magnetization yields a clover-like current pattern. 
In addition to that, we also find that a uniform charge current could induce a spin polarization around an impurity even above the spontaneous nucleation of the magnetization pattern. While this feature may be too weak to be easily observed, it is interesting to notice that currents could also influence and possibly deform the magnetically ordered pattern. 

With our theory we have illustrated that Ti-impurities naturally lead to inhomogeneous magnetic order. The complex structure of this order depends on the random positions of the Ti-ions. Since we are interested here on the possibility of nucleating magnetic order by Ti-doping, we leave the discussion of the random multi impurity system to future studies. Our study also does not intend to give a quantitative account of the phenomenon.

\section*{Acknowledgements}
We are very grateful to Jose L. Lado, Jun Goryo and Mark H. Fischer for many useful discussions. This work was financially supported by the Swiss National Science Foundation (SNSF) through Division II (No. 184739).


\appendix

\section{Effective field theory: Variational minimization} \label{sec:appendixSaddle}
In this section we elaborate on the details of the derivation of the effective gauge field theory used in Sec.~\ref{sec:gauge_formul}, starting with the functional integral form of the partition function of the microscopic model. 

We use the auxiliary fields $\phi_{ij} = (\phi_{ij}^{x}, \phi_{ij}^{y})$ to decouple the Coulomb interaction by means of a Hubbard-Stratonovich transformation, which results in the partition function including the integral over $ \phi_{ij} $, 
\begin{align}
    Z &= \int \mathcal{D}\phi \mathcal{D}C^{\dagger} \mathcal{D}C  \nonumber \\ 
    &\times \text{exp} \left[ - \int_{0}^{\beta} L'(\phi, C_{i j \sigma}^{\dagger}, C_{i j \sigma}) d\tau  \right], 
\end{align}
with 
\begin{align}
    L '= &\sum_{\bm{k},\bm{q},s} C^{\dagger}_{\bm{k_{-}}s} \left( \partial_{\tau} + \hat{H}_{\bm{k},s} \right) \delta_{\bm{q},0} C_{\bm{k_{+}}s} \nonumber \\[3mm]
    &+ \sum_{\bm{k},\bm{q},s,s'} C^{\dagger}_{\bm{k_{-}}s} \left(  \sum_{s'} \bm{A}_{-\bm{q}} \cdot \hat{\bm{I}}_{\bm{k}} \sigma^{0}_{s,s'} \right) C_{\bm{k_{+}}s'} \nonumber \\[3mm]
    &+ \sum_{\bm{k},\bm{q},s,s'} C^{\dagger}_{\bm{k_{-}}s} \left(  \sum_{s'} \bm{a}_{-\bm{q}} \cdot \hat{\bm{I}}_{\bm{k}} \sigma^{z}_{s,s'} \right) C_{\bm{k_{+}}s'} \nonumber \\[3mm]
    &+ \sum_{\bm{k},\bm{q},s} C^{\dagger}_{\bm{k_{-}}s} \left( V_{-\bm{q}} + s \hat{\phi}_{-\bm{q}}\right) C_{\bm{k_{+}}s} \nonumber \\[3mm]
& + \sum_{i,j} \sum_{\alpha, \alpha'} \phi_{ij}^{\alpha} U_{\alpha \alpha'}^{-1} \phi_{ij}^{\alpha'},\label{eqn:fullLagr}
\end{align}
where $ \hat{H}_{\bm{k},s} $ [Eq.~(\ref{H-matrix})] denotes the Hamiltonian and $ \hat{\bm{I}}_{\bm{k}} $ [Eq.~(\ref{eqn:current_matrix})] the current matrix. We use here the short-hand notation $\bm{k_{\pm}} = \bm{k} \pm \bm{q}/2$ and the inverse interaction matrix
\begin{align}
\hat{U}^{-1} = \frac{1}{U^2 - J^2} \begin{pmatrix} U & -J \\ -J & U \end{pmatrix} 
\end{align}
as well as  $ \hat{\phi}_{-\bm{q}} = {\rm diag}( \phi^{x}_{-\bm{q}}, \phi^{y}_{-\bm{q}}) $.
The potential of the impurities is given by $V_{-\bm{q}} = \sum_j V e^{i \bm{q}\cdot \bm{R}_{j}}$, where $\bm{R}_j$ are the impurity positions and $ V > 0 $. 

After integrating over the Fermionic degrees of freedom we end up with the effective action, which allows us to express the partition function as
\begin{align}
	Z &= \int \mathcal{D}\phi \; e^{-S_{\text{eff}}}, \label{eqn:effectact} \\[2mm] 
    &= \int \mathcal{D}\phi \Bigg \{ \exp \left[ -\sum_{\bm{q}, i \nu_l} \sum_{\alpha,\alpha'}  \phi_{\bm{q}}^{\alpha}(i \nu_l) U_{\alpha \alpha'}^{-1} \phi_{-\bm{q}}^{\alpha'} (-i\nu_l)
       \right] \nonumber \\[2mm]
&\times \exp \left[ \text{Tr} \ln \left( M \right)_{(\bm{k}_-, i\omega_n, s),(\bm{k}_+, i\omega_m, s')} \right] \Bigg \}, \label{eqn:actionexp}
\end{align}
with
\begin{align} 
&M_{(\bm{k}_{-}, i\omega_n,s),(\bm{k}_{+}, i\omega_m,s')} \nonumber \\[2mm] 
&= \Bigg[ (-i \omega_n + \hat{H}_{\bm{k},s}) \delta_{\omega_n, \omega_m} \delta_{\bm{q},0} +  \bm{A}_{-\bm{q}} \cdot \hat{\bm{I}}_{\bm{k}}
+s \bm{a}_{-\bm{q}} \cdot \hat{\bm{I}}_{\bm{k}} \nonumber \\ &\hspace{0.5cm}+ \frac{1}{\sqrt{\beta V}} V_{-\bm{q}} + \frac{s}{\sqrt{\beta V}} \hat{\phi}_{-\bm{q}} (i \omega_n -  i\omega_m) \Bigg] \sigma_{ss'}^0, \nonumber \\[2mm] 
&= - \hat{G}_{0,s}^{-1} \sigma_{ss'}^0 \Bigg[ 1 - \hat{G}_{0,s} \left( \bm{A}_{-\bm{q}} \cdot \hat{\bm{I}}_{\bm{k}} \right) - s \hat{G}_{0,s}  \left( \bm{a}_{-\bm{q}} \cdot \hat{\bm{I}}_{\bm{k}} \right) \nonumber \\ 
&\hspace{0.3cm}- \hat{G}_{0,s}  \frac{1}{\sqrt{\beta V}} V_{-\bm{q}} - \hat{G}_{0,s}  \frac{s}{\sqrt{\beta V}} \hat{\phi}_{-\bm{q}} (i \omega_n -  i\omega_m) \Bigg] .  \label{M-expansion} 
\end{align}
The bare Green's function denoted as $ \hat{G}_{0,s} $ is defined through
\begin{equation}
(i \omega_n - \hat{H}_{\bm{k},s})  \hat{G}_{0,s} (\bm{k}, i \omega_n) = \hat{\sigma}_0 ,
\end{equation}
with Fermionic (Bosonic) Matsubara frequencies, $ \omega_n = (2n+1) \pi /\beta $ ($ \nu_l = 2l \pi/\beta $). We may now expand the $ \ln (M) $ term using
\begin{align}
\ln (1-x) = -\sum_{n=1}^{\infty} \frac{x^n}{n} .
\end{align} 

Restricting ourselves to terms up to second order in the field $ \phi^{\alpha}_{\bm{q}} $, we obtain for the effective action the expression given in Eq.~\eqref{eqn:Seff},
which  can now be  used to analyze the magnetic stability of the system through variation with respect to $ \phi^{\alpha}_{\bm{q}} $ in the static case, 
\begin{align}
0 & = \left. \frac{\partial S_{\rm eff}^{(2)}}{\partial \phi_{\bm{q}}^{\alpha} } \right|_{i \nu_l=0} = \sum_{\alpha'} \left[ U_{\alpha \alpha'}^{-1} - \chi_{0}^{\alpha \alpha'} (\bm{q}) \right] \phi_{-\bm{q}}^{\alpha'} \\ \nonumber
& + V \sum_{j} e^{i \bm{q} \cdot \bm{R}_j} \sum_{\bm{q}', \alpha'}  e^{- i \bm{q}' \cdot \bm{R}_j} \Gamma_{\alpha \alpha'} (\bm{q}, \bm{q}') \phi_{-\bm{q}'}^{\alpha'}.
\end{align}
This corresponds to a linearized version of a Ginzburg-Landau theory. 

So far we have written the effective action in terms of the auxiliary fields $\phi_{\bm{q}}^{\alpha}$. Alternatively, we can also express $ S_{\rm eff}^{(2)} $ of Eq.~(\ref{eqn:Seff}) in terms of the magnetic moments $ m_{\bm{q}}^{\alpha}  = - \sum_{\alpha'} U_{\alpha \alpha'}^{-1} \phi_{\bm{q}}^{\alpha'} $, a relation we derive in App.~\ref{sec:phi_m_rel}, 
\begin{align}
\tilde{S}_{\rm eff}^{(2)} &=  \sum_{\alpha, \alpha'} \sum_{\bm{q}} m_{\bm{q}}^{\alpha} \tilde{A}_{\alpha \alpha'} (\bm{q})  m_{-\bm{q}}^{\alpha'} \nonumber \\
& + V \sum_{\alpha, \alpha'} \sum_{\bm{q},\bm{q},j} e^{i (\bm{q}-\bm{q}')\cdot \bm{R}_j} m_{\bm{q}}^{\alpha} \tilde{\Gamma}_{\alpha \alpha'} (\bm{q},\bm{q}')  m_{-\bm{q}'}^{\alpha'} ,
\end{align}
with
\begin{equation}
\hat{\tilde{A}}(\bm{q}) = \hat{U} [ \hat{U}^{-1} - \hat{\chi}_0 (\bm{q})] \hat{U} = [1-\hat{U} \hat{\chi}_0 (\bm{q})] \hat{U} \label{eqn:A_appendix}
\end{equation}
and
\begin{equation}
\hat{\tilde{\Gamma}}(\bm{q},\bm{q}' ) = \hat{U} \hat{\Gamma}(\bm{q},\bm{q}' )  \hat{U} .
\end{equation}
We may now use these expressions to estimate the parameters of the Ginzburg-Landau free energy functional given in Sec.~\ref{GL-Theory}. 
The symmetric matrix $ \hat{\tilde{A}}(\bm{q}) $ can be 
compared with $ \hat{A}(\bm{q}) $. The impurity term involves some spatial structure not present in the Ginzburg-Landau theory, where the impurity term is simply approximated by a delta-function,
\begin{align}
	 V \hat{\tilde{\Gamma}}(\bm{q},\bm{q}' ) \overset{\text{approx.}}{\longrightarrow} \gamma \hat{1}.
\end{align}
Most notably, we find that for positive $V$ the average of the left side is negative, which implies that $\gamma < 0$. 

The analysis of the instability by means of the field theory yields the result displayed in Fig.~\ref{fig:mf-gl-comp}. We observe qualitatively the same cross like structure as in the local mean field theory of Sect.~\ref{sec:impindpol}.

\section{The relation between $ \bm{m} $ and $\bm{\phi}$}\label{sec:phi_m_rel} 
We couple by a position dependent magnetic field to the electron spin specifying even the orbital,
\begin{align}
H_Z & = - \sum_{i,j,s,\alpha} H_{ij}^{\alpha} s c_{ij s \alpha}^{\dag}  c_{ij s \alpha} \\ \nonumber 
& = - \sum_{\bm{k},\bm{q},s,\alpha} H_{-\bm{q}}^{\alpha}  s c_{\bm{k}_- s \alpha}^{\dag}  c_{\bm{k}_+ s \alpha} ,
\end{align}
which included in the effective action [Eq.~\eqref{eqn:Seff}] leads to
\begin{align}
&S_{\rm eff}^{(2)}  =  \sum_{\bm{q}, i \nu_l} \sum_{\alpha,\alpha'}   \phi_{\bm{q}}^{\alpha}(i \nu_l)  U_{\alpha \alpha'}^{-1} \phi_{-\bm{q}}^{\alpha'} (-i\nu_l)  \nonumber \\ 
&- \sum_{\bm{q}, i \nu_l} \sum_{\alpha,\alpha'}  ( \phi_{\bm{q}}^{\alpha}(i \nu_l)- H_{\bm{q}}^{\alpha}) \chi_{0}^{\alpha \alpha'}(\bm{q}, i \nu_l)(\phi_{-\bm{q}}^{\alpha'} (-i\nu_l) - H_{-\bm{q}}^{\alpha'}) . 
\end{align}
Ignoring the impurities, the saddle point equation in the paramagnetic phase for the static field $ \phi_{\bm{q}}^{\alpha}  $ is 
\begin{align}
0 &=  \left. \frac{\partial S_{\rm eff}^{(2)}}{\partial \phi_{\bm{q}}^{\alpha}} \right|_{i \nu_l =0} \nonumber \\ 
&=  \sum_{\alpha'} \left\{ U_{\alpha \alpha'}^{-1} \phi_{-\bm{q}}^{\alpha'} - \chi_0^{\alpha \alpha'} (\bm{q},0) (\phi_{-\bm{q}}^{\alpha'} - H_{-\bm{q}}^{\alpha'}) \right\}, 
\label{saddle-H} 
\end{align}
and the magnetic moment $ m_{-\bm{q}}^{\alpha} $ can be defined through
\begin{align}
m_{-\bm{q}}^{\alpha}  &= -  \left. \frac{\partial S_{\rm eff}^{(2)}}{\partial H_{\bm{q}}^{\alpha}} \right|_{i \nu_l =0} \nonumber \\  &= - \sum_{\alpha'}  \chi_0^{\alpha \alpha'} (\bm{q},0)  (\phi_{-\bm{q}}^{\alpha'} - H_{-\bm{q}}^{\alpha'}) .
\label{mqa-definition}
\end{align}
Now combining Eqs.~(\ref{saddle-H}) and (\ref{mqa-definition}) we obtain for the magnetization
\begin{align}
m_{-\bm{q}}^{\alpha} & = \sum_{\alpha'} \left\{ \hat{\chi}_0(\bm{q},0) \left[ 1 - \hat{U} \hat{\chi}_0(\bm{q},0) \right]^{-1} \right\}_{\alpha \alpha'} H_{-\bm{q}}^{\alpha'}, \nonumber \\
& = \sum_{\alpha'} \chi^{\alpha \alpha'} (\bm{q}, 0) H_{-\bm{q}}^{\alpha'} ,
\label{m-rpa}
\end{align} 
which defines the renormalized susceptibility on the level of the random phase approximation. On the other hand, we can derive from the two Eqs.~(\ref{saddle-H}) and (\ref{mqa-definition}) the relation between $ m_{\bm{q}}^{\alpha} $ and $ \phi_{\bm{q}}^{\alpha} $,
 \begin{align}
m_{\bm{q}}^{\alpha}   = -\sum_{\alpha'} U_{\alpha \alpha'}^{-1}  \phi_{\bm{q}}^{\alpha'}  ,
\label{m-phi-rel}
\end{align}
which we used in the previous section.

\bibliography{references}

\end{document}